\pdfoutput=1
\documentclass[sigconf,nonacm]{acmart}
\AtBeginDocument{%
  }

\usepackage{enumitem}
\usepackage{multirow}
\usepackage{graphicx}
\usepackage{colortbl}
\usepackage{xspace}
\usepackage{amssymb}
\usepackage{varwidth}
\usepackage{booktabs}
\definecolor{bgcolor}{RGB}{242, 242, 242}
\usepackage{makecell}
\graphicspath{{./}}

\begin{document}

\title[DuoFormer]{DuoFormer: Parallel Shared and Task-Specific Modeling for Industrial Multi-Task Recommendation}
\author{Xiangming Li}
\authornote{These authors contributed equally to this work.}
\email{lixiangming.lxm@alibaba-inc.com}
\affiliation{%
  \institution{Alibaba Group}
  \city{Beijing}
  \country{China}
}

\author{Shijie Zhu}
\authornotemark[1]
\email{yangmu.zsj@alibaba-inc.com}
\affiliation{%
  \institution{Alibaba Group}
  \city{Beijing}
  \country{China}
}

\author{Kai Zhang}
\email{victorlanger.zk@alibaba-inc.com}
\affiliation{%
  \institution{Alibaba Group}
  \city{Beijing}
  \country{China}
}

\author{Bencheng Yan}
\email{bencheng.ybc@alibaba-inc.com}
\affiliation{%
  \institution{Alibaba Group}
  \city{Beijing}
  \country{China}
}

\author{Chuan Yu}
\email{yuchuan.yc@alibaba-inc.com}
\affiliation{%
  \institution{Alibaba Group}
  \city{Beijing}
  \country{China}
}

\author{Jian Xu}
\email{xiyu.xj@alibaba-inc.com}
\affiliation{%
  \institution{Alibaba Group}
  \city{Beijing}
  \country{China}
}

\renewcommand{\shortauthors}{Li et al.}

\begin{abstract}
Large-scale Web platforms model diverse user feedback and business objectives using multi-task recommender systems, which have recently achieved substantial performance gains by scaling model capacity. However, existing scaling approaches typically follow a \emph{sequential paradigm} where input features are first encoded into shared representations before task-specific modeling is introduced, with model scaling largely concentrated on expanding the shared encoder's capacity. Early feature interactions therefore remain task-agnostic even as shared capacity grows. Furthermore, these approaches rely on coarse-grained task modeling and task-invariant feature representations, constraining the growth of task-specific modeling capacity as models scale. To address these limitations, we introduce \textbf{DuoFormer}, a dual-channel architecture for scaling multi-task recommendation through a \emph{parallel paradigm} where fine-grained task modeling proceeds alongside shared feature learning from the earliest stages of interaction. Specifically, DuoFormer learns fine-grained task-specific representations through structured multi-view modeling and lightweight feature adaptation. To balance task specialization and cross-task knowledge transfer, we introduce role-aware communication that shares consolidated task information while preserving fine-grained task-specific representations. These mechanisms support efficient scaling within a unified backbone, where the shared channel reuses feature computation across tasks and the task-specific channel provides lightweight adaptation with limited additional overhead. Moreover, consensus--disagreement-aware gradient routing aligns optimization with the parallel dual-channel design without introducing additional inference overhead. Extensive offline experiments demonstrate strong overall performance and favorable scaling behavior, while online A/B tests demonstrate substantial gains in key business metrics.
\end{abstract}



\maketitle

\section{Introduction}

Large-scale Web platforms in e-commerce, content distribution, and online advertising rely on industrial recommender systems to deliver personalized recommendations from vast pools of items and content~\cite{covington2016youtube,ma2018esmm,li2026spectral}. To capture user preferences and accommodate diverse business requirements, these systems typically learn from multiple feedback signals and optimize multiple objectives within a unified framework~\cite{ma2018mmoe,tang2020ple}. For example, short-video platforms jointly predict user feedback such as likes, shares, comments, and video completion~\cite{gao2022kuairand, chen2025uniformer}. Consequently, multi-task learning has become a predominant modeling paradigm for recommender systems on these Web platforms~\cite{ma2018mmoe,tang2020ple}. Inspired by the success of model scaling in large language models (LLMs)~\cite{kaplan2020scaling,hoffmann2022training}, recent research on these recommender systems has increasingly explored how recommendation performance scales with model capacity and has yielded substantial gains~\cite{zhang2024wukong,zhai2024actions}.

\begin{figure}[t]
\centering
\includegraphics[width=1.0\linewidth]{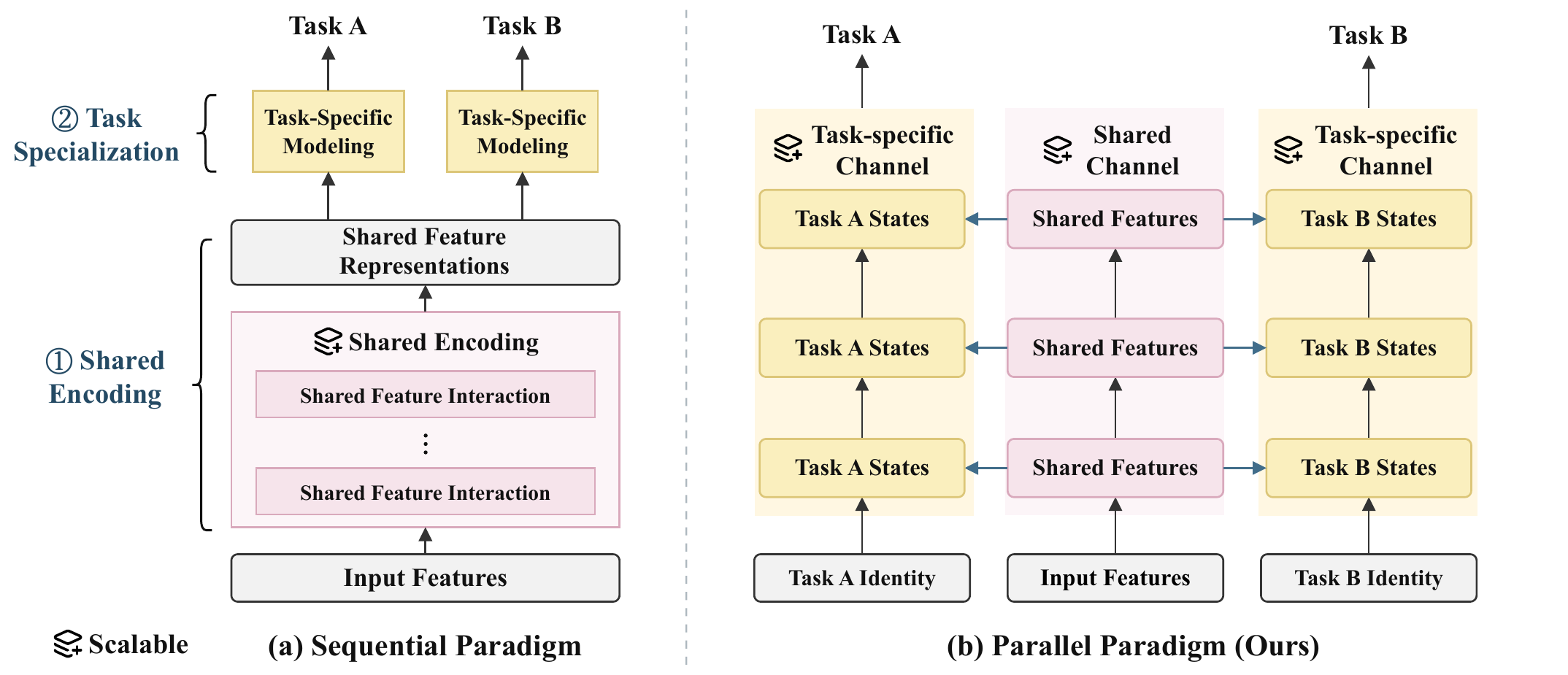}
\vspace{-2mm}
\caption{Comparison of conventional sequential paradigm and our proposed parallel paradigm.}
\label{fig:intro}
\vspace{-2mm}
\end{figure}

However, existing scaling approaches typically follow a \emph{sequential paradigm}, as illustrated in Figure~\ref{fig:intro}(a), where input features are first encoded into shared representations before task-specific modeling is introduced~\cite{zhu2025rankmixer, zhang2025onetrans,zhou2026tokenformer}. Within this framework, model scaling primarily expands the shared encoder's capacity, leaving two key limitations in task-specific modeling unresolved. First, task specialization begins only after multiple layers of shared feature interaction, leaving early interactions task-agnostic even as shared capacity grows. All tasks therefore rely on a common, highly abstracted feature view from which fine-grained task-relevant information can be difficult to disentangle. Subsequent modules primarily select and combine these shared representations without explicitly adapting individual feature representations to task-specific requirements. Second, task representations remain coarse-grained as models scale, with diverse preference factors and behavioral patterns underlying each task often aggregated into a single tower state or task token \cite{chen2025uniformer, jiang2026tokenmixerlarge, zhang2025onetrans}. Their distinct feature dependencies and individual contributions remain implicit, constraining the growth of task-specific modeling capacity despite increases in overall model size.

To overcome these limitations, we advocate a \emph{parallel paradigm} for scaling multi-task recommendation, in which fine-grained task-specific modeling proceeds alongside shared representation learning from the earliest stages of feature interaction, as illustrated in Figure~\ref{fig:intro}(b). The parallel organization of scalable shared and task-specific channels allows each task to shape feature retrieval and interpretation as shared representations evolve, enabling increased model capacity to strengthen both shared feature learning and fine-grained task-specific modeling.

However, realizing this parallel paradigm poses three key challenges. First, \emph{how to learn fine-grained task representations before high-level shared representations are available?} Without high-level shared representations, each task-specific channel has limited contextual information about dependencies among features. It must therefore jointly learn how to combine fine-grained feature information into a coherent task representation and how to interpret individual features according to task-specific requirements. Second, \emph{how to enable cross-task knowledge transfer while preserving task specialization?} Isolated task channels limit knowledge sharing across related tasks, whereas unrestricted interaction can propagate task-private information and blur task-specific distinctions~\cite{ma2018mmoe,tang2020ple}. Third, \emph{how to jointly scale shared and task-specific channels without excessive inference overhead?} A straightforward implementation of the parallel paradigm would introduce separate task-specific processing paths starting at the early stages of feature interaction, duplicating substantial downstream computation. This redundancy becomes increasingly costly as model capacity grows, making it difficult to meet the inference latency and resource constraints of industrial recommender systems~\cite{chen2025uniformer,zhang2026smes}.

To address these challenges, we propose \textbf{DuoFormer}, a dual-channel architecture for scaling industrial multi-task recommendation that integrates shared feature interaction, fine-grained task-specific modeling, and selective cross-task communication within a unified interaction module. Specifically, DuoFormer learns fine-grained task representations during early feature interaction through structured multi-view task modeling and lightweight feature adaptation. The multi-view structure follows a hub--facet hierarchy that separates information extraction from task-level integration. Multiple facet tokens capture fine-grained information from distinct task-relevant perspectives, while a hub token coordinates these views and integrates them into a coherent task representation. Lightweight feature adaptation uses task-conditioned low-rank transformations to tailor shared feature representations for individual tasks, refining how each task retrieves and interprets feature information. To preserve fine-grained task specialization while enabling cross-task knowledge transfer, we introduce role-aware communication that keeps facet representations local to each task and restricts cross-task exchange to consolidated hub representations. This enables complementary knowledge sharing while protecting fine-grained task representations from direct cross-task interference. To support efficient scaling of both shared and task-specific channels under industrial serving constraints, we integrate them within a unified interaction backbone that reuses shared feature computation across tasks and limits task-specific overhead through lightweight adaptation. Furthermore, we introduce consensus--disagreement-aware gradient routing, which uses the distinct update paths of the parallel design to preserve compatible shared updates and emphasize task-specific adaptation when gradients conflict, without adding inference overhead. Our main contributions are summarized as follows:

\begin{itemize}[
    topsep=2pt,
    itemsep=2pt,
    parsep=0pt,
    partopsep=0pt
]
    \item We advocate a shift from the sequential paradigm to a parallel paradigm for scaling multi-task recommendation, in which fine-grained task-specific modeling proceeds alongside shared representation learning from the earliest stages of feature interaction.
    
    \item We propose DuoFormer, a scalable dual-channel architecture that integrates hub--facet multi-view modeling, low-rank feature adaptation, and role-aware communication within a shared interaction backbone, enabling fine-grained task specialization and cross-task knowledge sharing with limited additional computation.
    
    \item We conduct extensive experiments and in-depth analyses that demonstrate DuoFormer's strong recommendation performance and favorable scaling behavior, while validating the contributions of its key components. Online A/B tests further demonstrate DuoFormer's substantial business impact, with CVR and GMV increasing by 1.40\% and 1.41\%, respectively.
\end{itemize}

\section{Preliminaries}
\label{sec:preliminaries}

Let $\mathbf{X}$ denote the input features of a sample and $T$ the number of prediction tasks. A feature embedding module maps $\mathbf{X}$ to a collection of dense feature embeddings, denoted by $\mathbf{E}_s = F_{\mathrm{embed}}(\mathbf{X})$. For task $t$, we denote its prediction and training loss by $\widehat{y}_t$ and $\mathcal{L}_t$, respectively.

Representative scaling approaches for industrial multi-task recommendation follow a sequential design, first learning shared feature representations and then performing task-specific modeling to produce predictions~\cite{zhu2025rankmixer, jiang2026tokenmixerlarge,huang2026mixformer, zhang2025onetrans,zhou2026tokenformer, chen2025uniformer}. At a high level, this workflow can be expressed as:
\begin{equation}
    \begin{aligned}
        \mathbf{R}_s
        &= F_{\mathrm{shared}}(\mathbf{E}_s), \\
        \mathbf{R}_t
        &= F_t(\mathbf{R}_s),
        \qquad
        \widehat{y}_t = P_t(\mathbf{R}_t),
        \quad t=1,\ldots,T.
    \end{aligned}
    \label{eq:sequential_mtl}
\end{equation}
Here, $F_{\mathrm{shared}}$ denotes the shared feature encoder, while $F_t$ and $P_t$ denote the task-specific modeling module and prediction head for task $t$, respectively. The representations $\mathbf{R}_s$ and $\mathbf{R}_t$ are the outputs of shared encoding and task-specific modeling, respectively.

We instead formulate a parallel paradigm for scaling multi-task recommendation, in which fine-grained task-specific modeling proceeds alongside shared representation learning from the earliest stages of feature interaction. Let $\mathbf{E}_{\mathrm{task}} = [\mathbf{e}_1;\ldots;\mathbf{e}_T]$ collect the initial task representations, where $\mathbf{e}_t$ encodes task-specific attributes such as task ID and objective type. The parallel formulation is:
\begin{equation}
    \begin{aligned}
        &\mathbf{R}_{\mathrm{tasks}}
        = F_{\mathrm{parallel}}
        \left(\mathbf{E}_s,\mathbf{E}_{\mathrm{task}}\right), \\
        &\widehat{y}_t
        = P_t(\mathbf{R}_t),
        \quad t=1,\ldots,T.
    \end{aligned}
    \label{eq:parallel_mtl}
\end{equation}
Here, $\mathbf{R}_{\mathrm{tasks}} = [\mathbf{R}_1;\ldots;\mathbf{R}_T]$ collects the task-specific outputs supplied to the prediction heads. The mapping $F_{\mathrm{parallel}}$ jointly develops shared feature representations and task-specific states through shared feature interaction, task-dependent feature interpretation, and cross-task communication. Task-specific states are therefore refined as shared representations evolve, enabling task-dependent feature retrieval and interpretation during the interaction process.

\begin{figure*}[t]
\centering
\includegraphics[width=0.96\linewidth]{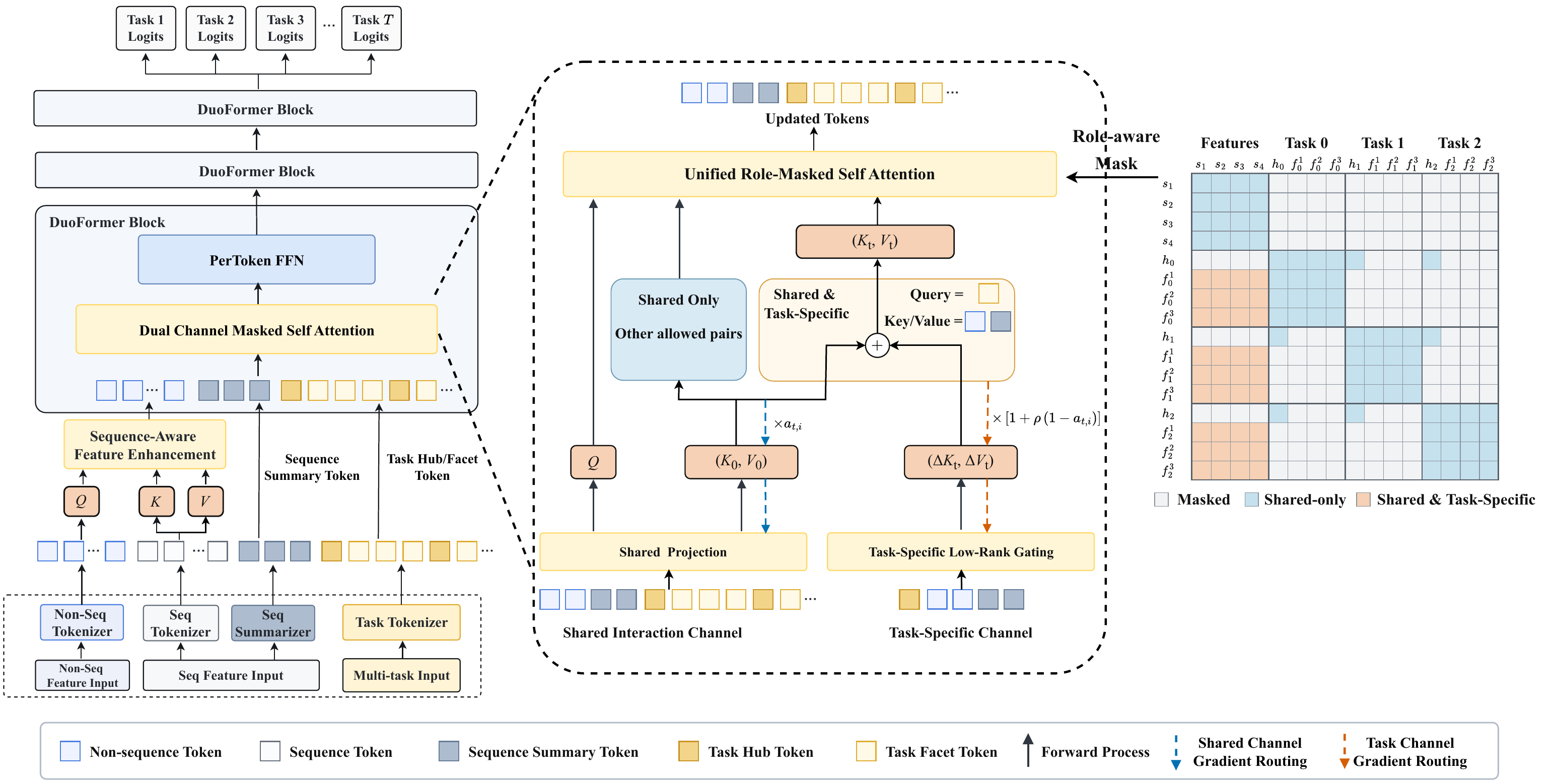}
\vspace{-3mm}
\caption{The overall framework of the proposed DuoFormer.}
\label{fig:overall}
\vspace{-2mm}
\end{figure*}

\section{Methodology}
\label{sec:methodology}

As illustrated in Figure~\ref{fig:overall}, DuoFormer realizes the parallel paradigm for scaling multi-task recommendation through a unified dual-channel backbone that integrates shared feature learning, multi-view task modeling, and lightweight task-dependent feature adaptation from the earliest stages of feature interaction. Specifically, we first construct heterogeneous feature tokens and hub--facet task tokens, then enrich non-sequential features with behavioral context. Within a unified interaction backbone, the shared channel integrates shared feature interaction, hub--facet task modeling, and cross-task communication through role-aware masked attention. In parallel, the task-specific channel tailors shared feature representations to individual tasks through lightweight low-rank adaptations, enabling task-dependent feature interpretation with limited additional computational cost. Finally, to align optimization with the distinct roles of the two channels, we introduce consensus--disagreement-aware gradient routing to regulate shared and task-specific updates without adding inference overhead.

\subsection{Heterogeneous Tokenization}

Industrial recommender systems operate on heterogeneous input features with distinct structures. Following common practice in scalable recommendation models~\cite{zhu2025rankmixer,ha2026unimixer,zhang2025onetrans}, we map the input feature embeddings $\mathbf{E}_s$ into a common token space for subsequent interaction. Specifically, we construct non-sequential feature tokens and sequence element and summary tokens, then enrich the non-sequential tokens with behavioral context. We also initialize hub--facet task tokens for fine-grained task modeling.

\subsubsection{Non-Sequential Feature Tokenization}

Non-sequential inputs comprise categorical and numerical attributes describing the user, candidate item, and request context. We organize these fields into $N$ semantic groups, with $n_i$ fields in group $i$. Let $\mathbf{e}_{i,j}$ denote the dense representation of the $j$-th field in group $i$. We concatenate the field representations within each group, project them into a common $d$-dimensional token space, and add a non-sequential type embedding before layer normalization.
\begin{equation}
    \mathbf{s}_i^{\mathrm{nonseq}}
    =
    \operatorname{LN}
    \left(
        \phi_i
        \left(
            \operatorname{concat}
            \left(
                \mathbf{e}_{i,1},
                \ldots,
                \mathbf{e}_{i,n_i}
            \right)
        \right)
        +
        \mathbf{e}^{\mathrm{type}}_{\mathrm{nonseq}}
    \right),
\end{equation}
where $\phi_i(\cdot)$ denotes a group-specific projection, $\mathbf{e}^{\mathrm{type}}_{\mathrm{nonseq}} \in \mathbb{R}^{d}$ is a type embedding shared by all non-sequential tokens, and $\operatorname{LN}$ denotes layer normalization. Stacking the resulting group tokens yields:
\begin{equation}
    \mathbf{S}^{\mathrm{nonseq}}
    =
    [\mathbf{s}_1^{\mathrm{nonseq}};
     \ldots;
     \mathbf{s}_N^{\mathrm{nonseq}}]
    \in \mathbb{R}^{N\times d}.
\end{equation}

\subsubsection{Sequential Behavior Tokenization}

Behavior sequences reveal user interests through both individual interactions and collective behavioral patterns~\cite{zhou2018din,zhang2025onetrans}. We therefore represent each sequence at two complementary levels. Element tokens retain fine-grained behavioral information for sequence-aware feature enhancement, while summary tokens provide compact sequence-level representations for the unified interaction backbone.

Let $M$ denote the number of behavior sequences and $L_m$ the length of sequence $m$. The $j$-th element of sequence $m$ contains $c_{m,j}$ side-information fields, with $\mathbf{e}_{m,j,k}$ denoting the dense representation of its $k$-th field. We construct an element token by concatenating these field representations, projecting them into the common token space, and adding a sequential type embedding before layer normalization.
\begin{equation}
    \mathbf{b}_{m,j}
    =
    \operatorname{LN}
    \left(
        \psi_m
        \left(
            \operatorname{concat}
            \left(
                \mathbf{e}_{m,j,1},
                \ldots,
                \mathbf{e}_{m,j,c_{m,j}}
            \right)
        \right)
        +
        \mathbf{e}^{\mathrm{type}}_{\mathrm{seq}}
    \right),
\end{equation}
where $\psi_m(\cdot)$ is a sequence-specific projection into the common $d$-dimensional token space, and $\mathbf{e}^{\mathrm{type}}_{\mathrm{seq}} \in \mathbb{R}^{d}$ is a type embedding shared by all sequence element tokens. Stacking the element tokens of sequence $m$ yields:
\begin{equation}
    \mathbf{B}_m
    =
    [\mathbf{b}_{m,1};\ldots;\mathbf{b}_{m,L_m}]
    \in \mathbb{R}^{L_m\times d}.
\end{equation}

To capture sequence-level interests, we aggregate the element tokens into a summary token using an aggregation operator $\operatorname{summarize}_m$. This operator can be instantiated with mean pooling, attention pooling, or a Transformer-based encoder. A padding mask $\mathbf{m}_m$ excludes padded positions from the aggregation. Stacking the summary tokens from all $M$ behavior sequences yields:
\begin{equation}
    \begin{aligned}
        \mathbf{s}_m^{\mathrm{seq}}
        &=
        \operatorname{summarize}_m
        (\mathbf{B}_m,\mathbf{m}_m), \\
        \mathbf{S}^{\mathrm{seq}}
        &=
        [\mathbf{s}_1^{\mathrm{seq}};
         \ldots;
         \mathbf{s}_M^{\mathrm{seq}}]
        \in \mathbb{R}^{M\times d}.
    \end{aligned}
\end{equation}

\subsubsection{Sequence-Aware Feature Enhancement}

We further enrich non-sequential feature tokens with behavioral context through a single cross-attention layer over sequence element tokens. The retrieved behavioral information is incorporated into each feature token through a residual connection:
\begin{equation}
    \widetilde{\mathbf{S}}^{\mathrm{nonseq}}
    =
    \mathbf{S}^{\mathrm{nonseq}}
    +
    \operatorname{CA}
    \left(
        \operatorname{LN}(\mathbf{S}^{\mathrm{nonseq}}),
        \mathbf{K}_{\mathrm{seq}},
        \mathbf{V}_{\mathrm{seq}};
        \mathbf{m}
    \right).
\end{equation}
Here, $\mathbf{K}_{\mathrm{seq}}$ and $\mathbf{V}_{\mathrm{seq}}$ are obtained by projecting the concatenated sequence element tokens $[\mathbf{B}_1;\ldots;\mathbf{B}_M]$, $\mathbf{m}$ is the corresponding padding mask, and $\operatorname{CA}(\cdot)$ denotes multi-head cross-attention.

\subsubsection{Hierarchical Task Tokenization}

A prediction objective in industrial recommendation may depend on multiple behavioral factors with distinct feature dependencies~\cite{chen2025mal}. Aggregating these factors into a single task state leaves their different information needs implicit. We therefore construct a hub--facet hierarchy for each task, providing multiple modeling views while maintaining a coherent task-level representation.

For task $t$, we use the initial task representation defined in Section~\ref{sec:preliminaries} as its hub token, i.e., $\mathbf{h}^{(0)}_t = \mathbf{e}_t \in \mathbb{R}^{d}$. We then initialize $P$ facet tokens by adding distinct learnable view embeddings to the hub representation:
\begin{equation}
    \mathbf{f}_{t,p}^{(0)}
    =
    \operatorname{LN}
    \left(
        \mathbf{h}_t^{(0)}
        +
        \mathbf{e}^{\mathrm{facet}}_{p}
    \right),
    \qquad p=1,\ldots,P,
\end{equation}
where $\mathbf{e}^{\mathrm{facet}}_{p} \in \mathbb{R}^{d}$ denotes the learnable embedding for the $p$-th modeling view. This initialization conditions all facets on the same task identity while providing distinct starting points for learning task-relevant perspectives. Stacking the hub and facet tokens across all tasks yields $\mathbf{H}^{(0)} \in \mathbb{R}^{T\times d}$ and $\mathbf{F}^{(0)} \in \mathbb{R}^{TP\times d}$, respectively.

\subsection{Parallel Dual-Channel Interaction}

To enable fine-grained task specialization during early feature interaction, we introduce a dual-channel architecture that learns shared and task-specific representations in parallel. The shared channel integrates shared feature interaction, hub--facet task modeling, and role-aware cross-task communication, while the task-specific channel tailors feature retrieval and interpretation to individual tasks through lightweight low-rank adaptation.

\subsubsection{Shared Interaction Channel with Role-Aware Masking}

At layer $\ell$, it operates on a heterogeneous token sequence defined as
\begin{equation}
\begin{aligned}
    \mathbf{Z}^{(\ell)}
    &=
    [\mathbf{S}^{(\ell)};
     \mathbf{H}^{(\ell)};
     \mathbf{F}^{(\ell)}], \\[3pt]
    \mathbf{S}^{(0)}
    &=
    [\widetilde{\mathbf{S}}^{\mathrm{nonseq}};
     \mathbf{S}^{\mathrm{seq}}],
\end{aligned}
\end{equation}
where $\mathbf{S}^{(\ell)}$, $\mathbf{H}^{(\ell)}$, and $\mathbf{F}^{(\ell)}$ denote shared feature tokens, task hub tokens, and task facet tokens, respectively.

Allowing all tokens to attend to one another can entangle reusable feature information with task-conditioned signals and expose task-private representations to direct cross-task interference. We therefore introduce role-aware masked self-attention that blocks direct information flow from task tokens to shared feature tokens and restricts cross-task communication to hub tokens carrying consolidated task-level information. The resulting attention operation is defined as follows:
\begin{equation}
\begin{gathered}
    \widehat{\mathbf{Z}}^{(\ell)}
    =
    \operatorname{MaskedSelfAttn}
    \left(
        \mathbf{Z}^{(\ell)};\mathbf{M}
    \right),
    \\[6pt]
    \mathbf{M}
    =
    \begin{array}{c|ccc}
        & \mathbf{S}^{(\ell)}
        & \mathbf{H}^{(\ell)}
        & \mathbf{F}^{(\ell)}
        \\ \hline
        \mathbf{S}^{(\ell)}
        & \mathbf{0}
        & -\infty
        & -\infty
        \\[3pt]
        \mathbf{H}^{(\ell)}
        & -\infty
        & \mathbf{0}
        & \mathbf{M}_{\mathrm{intra}}
        \\[3pt]
        \mathbf{F}^{(\ell)}
        & \mathbf{0}
        & \mathbf{M}_{\mathrm{intra}}
        & \mathbf{M}_{\mathrm{intra}}
    \end{array},
    \\[6pt]
    \mathbf{M}_{\mathrm{intra}}(\mathbf{q},\mathbf{k})
    =
    \begin{cases}
        0,       & \operatorname{task}(\mathbf{q})=\operatorname{task}(\mathbf{k}), \\
        -\infty, & \text{otherwise}.
    \end{cases}
\end{gathered}
\end{equation}
The row and column blocks of $\mathbf{M}$ correspond to query and key--value token groups, respectively. Entries of $0$ allow attention, while entries of $-\infty$ block it. Each $\mathbf{M}_{\mathrm{intra}}$ block permits attention only between tokens belonging to the same task, where $\operatorname{task}(\cdot)$ denotes the task index of a hub or facet token.

By structuring information flow according to token roles, the mask unifies four complementary interaction routes within a single self-attention operation, as detailed below. For clarity, we omit layer indices and use $\mathbf{F}_t$ to denote the facet tokens of task $t$. Arrows indicate information flow from key--value tokens to query tokens.

\textbf{Shared Feature Interaction.} To preserve a reusable feature basis across tasks, each shared feature token attends exclusively to the full shared feature set through the $\mathbf{S}\leftrightarrow\mathbf{S}$ route. This attention configuration captures dependencies across heterogeneous feature sources while reducing entanglement with task-conditioned signals.

\textbf{Facet-to-Feature Retrieval.} To accommodate diverse information needs within each task, every facet directly queries the full shared feature set through the $\mathbf{S}\rightarrow\mathbf{F}_t$ route. Facet-specific queries selectively retrieve and aggregate feature information to form distinct task-relevant views. Maintaining these views in distinct facet representations separates multi-view feature extraction from task-level consolidation and provides fine-grained information for subsequent intra-task coordination.

\textbf{Intra-Task Hub--Facet Coordination.} To refine and integrate multiple task views, facets within each task interact with one another and exchange information bidirectionally with their hub through $\mathbf{F}_t\leftrightarrow\mathbf{F}_t$ and $\mathbf{F}_t\leftrightarrow\mathbf{h}_t$. Facet-to-facet interactions support mutual refinement across views, while the hub consolidates their information into a coherent task representation and provides task-level context to guide further facet refinement.

\textbf{Inter-Task Hub Communication.} To share complementary knowledge while preserving task-specific distinctions, hubs serve as the sole interface for cross-task communication. By default, each hub attends to all hubs through the $\mathbf{H}\leftrightarrow\mathbf{H}$ route. When tasks follow an explicit dependency order, a causal mask instead allows each hub to attend only to itself and preceding hubs, enabling downstream tasks to incorporate upstream information while blocking reverse communication. Both configurations exchange consolidated task-level information through hubs while keeping fine-grained facet representations local to each task.

\subsubsection{Low-Rank Task-Specific Channel}

The shared channel enables task-dependent retrieval through distinct facet queries, but provides the same feature representations to all tasks. The same feature can indicate different user preferences or behavioral tendencies across tasks~\cite{su2024stem}, requiring explicit adaptation of its representation to each task's predictive requirements. We therefore introduce a lightweight task-specific channel that adapts shared feature representations to individual tasks, allowing each task to form its own relevance assessment and interpretation of the same feature with limited additional computational cost.

Specifically, we refine feature keys and values to adapt feature matching and content representation to each task, respectively. To control these refinements according to both task requirements and feature characteristics, we generate separate key and value gates from the initial task hub $\mathbf{h}^{(0)}_t$ and the shared feature token $\mathbf{s}_i$ using shared gating projections:
\begin{equation}
    \boldsymbol{\gamma}^{X}_{t,i}
    =
    \tanh
    \left(
        \sigma
        \left(
            \mathbf{h}^{(0)}_{t}\mathbf{W}^{X}_{H}
            +
            \mathbf{s}_{i}\mathbf{W}^{X}_{S}
        \right)
        \mathbf{W}^{X}_{g}
    \right),
    \qquad X\in\{K,V\}.
\end{equation}
Here, $\boldsymbol{\gamma}^{K}_{t,i}$ and $\boldsymbol{\gamma}^{V}_{t,i}$ denote the $r$-dimensional key and value gating vectors for task $t$ and shared feature $i$, respectively. The matrices $\mathbf{W}^{X}_{H},\mathbf{W}^{X}_{S}\in\mathbb{R}^{d\times d_g}$ project the task and feature representations into a common hidden space, and $\mathbf{W}^{X}_{g}\in\mathbb{R}^{d_g\times r}$ maps their nonlinear combination to an $r$-dimensional gating vector. The function $\sigma(\cdot)$ denotes a nonlinear activation, while $\tanh(\cdot)$ bounds the gating coefficients.

We then use these gates to modulate shared low-rank feature projections and construct task-specific key and value increments:
\begin{equation}
\left\{
\begin{aligned}
    \Delta\mathbf{k}_{t,i}
    &=
    \frac{\alpha}{r}
    \left[
        \left(
            \mathbf{s}_{i}\mathbf{A}_{K}
        \right)
        \odot
        \boldsymbol{\gamma}^{K}_{t,i}
    \right]
    \mathbf{B}_{K},
    \\[3pt]
    \Delta\mathbf{v}_{t,i}
    &=
    \frac{\alpha}{r}
    \left[
        \left(
            \mathbf{s}_{i}\mathbf{A}_{V}
        \right)
        \odot
        \boldsymbol{\gamma}^{V}_{t,i}
    \right]
    \mathbf{B}_{V}.
\end{aligned}
\right.
\end{equation}
Here, $\mathbf{A}_{K},\mathbf{A}_{V}\in\mathbb{R}^{d\times r}$ project shared features into compact adaptation spaces, while $\mathbf{B}_{K},\mathbf{B}_{V}\in\mathbb{R}^{r\times d_h}$ map the gated representations into the attention-head space, where $d_h$ denotes the dimension of each attention head. The rank $r$ controls adaptation capacity, $\alpha$ controls the update scale, and $\odot$ denotes element-wise multiplication. The key increment $\Delta\mathbf{k}_{t,i}$ adjusts the feature's compatibility with task-specific facet queries, while the value increment $\Delta\mathbf{v}_{t,i}$ adapts the information contributed by that feature. With $r\ll d$, all tasks reuse the same low-rank projection bases, while lightweight gates capture task--feature differences. This provides fine-grained feature adaptation without introducing separate dense projections for individual tasks.

We augment shared feature projections with lightweight task-dependent increments, enabling distinct feature interpretations within a single attention backbone shared across tasks. These increments apply exclusively to facet-to-feature retrieval along the $\mathbf{S}\rightarrow\mathbf{F}_t$ route, incorporating adapted feature information into task-local facet representations while preserving a single set of shared feature tokens. This allows shared feature encoding to scale without replicating its computation for each task. For task $t$, the adapted feature projections are given by:
\begin{equation}
\left\{
\begin{aligned}
    \mathbf{k}_{t,i}
    &=
    \mathbf{k}_{0,i}
    +
    \Delta\mathbf{k}_{t,i},
    \\[3pt]
    \mathbf{v}_{t,i}
    &=
    \mathbf{v}_{0,i}
    +
    \Delta\mathbf{v}_{t,i},
\end{aligned}
\right.
\end{equation}
where $\mathbf{k}_{0,i}$ and
$\mathbf{v}_{0,i}$ are the task-invariant feature projections provided by the shared channel.

The query $\mathbf{q}_{t,p}$ of facet $p$ in task $t$ scores shared feature $i$ using both the shared key and its task-specific increment:
\begin{equation}
    \ell^{S}_{t,p,i}
    =
    \underbrace{
        \frac{
            \mathbf{q}_{t,p}\mathbf{k}_{0,i}^{\top}
        }{\sqrt{d_h}}
    }_{\text{shared matching}}
    +
    \underbrace{
        \frac{
            \mathbf{q}_{t,p}\Delta\mathbf{k}_{t,i}^{\top}
        }{\sqrt{d_h}}
    }_{\text{task-specific refinement}}.
\end{equation}
For each facet query, a single softmax normalizes the logits over all tokens permitted by the mask. Let $\pi_{t,p,i}$ denote the attention weight assigned to shared feature $i$. The shared-feature contribution to the facet output is then given by:
\begin{equation}
    \mathbf{o}^{S}_{t,p}
    =
    \sum_i
    \pi_{t,p,i}
    \bigl(
        \underbrace{
            \mathbf{v}_{0,i}
        }_{\text{shared content}}
        +
        \underbrace{
            \Delta\mathbf{v}_{t,i}
        }_{\text{task-specific content}}
    \bigr).
\end{equation}

\subsubsection{Layer Updates and Task Prediction}

The dual-channel masked self-attention yields the output $\widehat{\mathbf{Z}}^{(\ell)}$. The token representations are then updated through residual connections and a feed-forward sublayer:
\begin{equation}
    \begin{aligned}
        \widetilde{\mathbf{Z}}^{(\ell)}
        &=
        \mathbf{Z}^{(\ell)}
        +
        \widehat{\mathbf{Z}}^{(\ell)}, \\
        \mathbf{Z}^{(\ell+1)}
        &=
        \widetilde{\mathbf{Z}}^{(\ell)}
        +
        \operatorname{FFN}
        \left(
            \operatorname{LN}
            \left(\widetilde{\mathbf{Z}}^{(\ell)}\right)
        \right).
    \end{aligned}
\end{equation}
Here, $\operatorname{FFN}$ denotes a position-wise feed-forward network, and $\ell=0,\ldots,L-1$ indexes the interaction layers. Increasing the backbone depth $L$ or token dimension $d$ expands the capacity of both shared feature learning and hub--facet task modeling within the same interaction architecture.

After the final interaction layer, we concatenate the hub and facet tokens of each task and use $P_t$ to map the resulting representation to its final prediction score:
\begin{equation}
    \begin{aligned}
        \mathbf{R}_t
        &=
        \operatorname{concat}
        \left(
            \mathbf{h}_t^{(L)},
            \mathbf{f}_{t,1}^{(L)},
            \ldots,
            \mathbf{f}_{t,P}^{(L)}
        \right), \\
        \widehat{y}_t
        &=
        P_t(\mathbf{R}_t),
        \qquad t=1,\ldots,T.
    \end{aligned}
\end{equation}
We train DuoFormer by minimizing a weighted sum of the losses across all $T$ tasks:
\begin{equation}
    \mathcal{L}
    =
    \sum_{t=1}^{T}\lambda_t\mathcal{L}_t,
\end{equation}
where $\mathcal{L}_t=\operatorname{loss}_t(y_t,\widehat{y}_t)$ denotes the loss for task $t$, $y_t$ is the corresponding ground-truth label, and $\lambda_t$ controls the contribution of each task to the overall objective.

\subsection{Consensus--Disagreement Gradient Routing}
Different task objectives can induce conflicting updates to the same shared feature representations. Existing methods for mitigating gradient conflicts primarily modify task gradients to coordinate shared updates~\cite{yu2020gradient,liu2021conflict}. However, their focus on coordinating shared updates does not explicitly exploit the task-specific supervisory information that gradient disagreement may reflect. Our parallel dual-channel design provides complementary paths for shared learning and task-specific adaptation, which allows task gradients to be routed according to agreement across tasks. Building on this property, we introduce Consensus--Disagreement Gradient Routing (CDGR), which preserves compatible shared updates and uses feature-level gradient disagreement to attenuate shared-path gradients while strengthening adaptation-path gradients.

To assess inter-task agreement at the feature level, we construct a gradient signature for task $t$ by concatenating the gradients of $\mathcal{L}_t$ with respect to the shared key and value projections of feature $i$:

\begin{equation}
    \mathbf{g}_{t,i}
    =
    \operatorname{concat}
    \left(
        \frac{\partial\mathcal{L}_{t}}
             {\partial\mathbf{k}_{0,i}},
        \frac{\partial\mathcal{L}_{t}}
             {\partial\mathbf{v}_{0,i}}
    \right).
\end{equation}

The gradients with respect to the shared key and value projections characterize task $t$'s preferred updates to feature matching and information representation, respectively. To reduce sensitivity to gradient magnitude and transient estimation noise~\cite{fernando2023mitigating}, we normalize each signature and maintain an exponential moving average:

\begin{equation}
    \bar{\mathbf{g}}_{t,i}
    \leftarrow
    \beta\bar{\mathbf{g}}_{t,i}
    +
    (1-\beta)
    \frac{\mathbf{g}_{t,i}}
         {\|\mathbf{g}_{t,i}\|_2+\epsilon},
\end{equation}
where $\beta\in[0,1)$ controls temporal smoothing and $\epsilon>0$ ensures numerical stability. Routing coefficients for the current training step are computed from the previous step's moving averages and held fixed during backpropagation. The newly collected gradients update the moving averages for the next training step.

For each task $t$, we construct a feature-specific consensus reference by aggregating the smoothed gradient signatures of the other tasks:
\begin{equation}
    \mathbf{c}_{-t,i}
    =
    \sum_{u\neq t}
    \omega_{u,i}\bar{\mathbf{g}}_{u,i},
\end{equation}
Here, the subscript $-t$ indicates that the consensus is constructed from all tasks other than task $t$. The weight $\omega_{u,i}$ controls the contribution of task $u$ to the consensus for feature $i$. We use equal weights by default, setting $\omega_{u,i}=1$. We then quantify directional disagreement between task $t$'s smoothed gradient signature and the consensus reference for the same feature:
\begin{equation}
    d_{t,i}
    =
    \operatorname{ReLU}
    \left(
        -\cos
        \left(
            \bar{\mathbf{g}}_{t,i},
            \mathbf{c}_{-t,i}
        \right)
    \right).
\end{equation}
The score is zero for non-opposing directions and increases with directional conflict. We map it to a continuous routing coefficient:
\begin{equation}
    a_{t,i}
    =
    \exp
    \left(
        -\frac{d_{t,i}}{\tau}
    \right),
\end{equation}
where $\tau>0$ controls sensitivity to disagreement. In the absence of directional conflict, $a_{t,i}=1$. Stronger disagreement yields a smaller coefficient, which we use to attenuate shared-path gradients and increase the emphasis on task-specific adaptation.

To apply this routing signal without changing the forward representations, we introduce the gradient-routing operator:
\begin{equation}
    \mathcal{R}(\mathbf{x},a)
    =
    \operatorname{sg}(\mathbf{x})
    +
    a
    \left(
        \mathbf{x}-\operatorname{sg}(\mathbf{x})
    \right),
\end{equation}
where $\operatorname{sg}(\cdot)$ denotes the stop-gradient operator. For a fixed routing coefficient $a$, $\mathcal{R}$ preserves the forward value of $\mathbf{x}$ while scaling its backward gradient by $a$. We use this operator to reduce gradients through the shared projections under disagreement while increasing the relative emphasis on task-specific adaptation:
\begin{equation}
\left\{
\begin{aligned}
    \mathbf{k}_{t,i}
    &=
    \mathcal{R}
    \left(
        \mathbf{k}_{0,i},a_{t,i}
    \right)
    +
    \mathcal{R}
    \left(
        \Delta\mathbf{k}_{t,i},
        1+\rho(1-a_{t,i})
    \right),
    \\[3pt]
    \mathbf{v}_{t,i}
    &=
    \mathcal{R}
    \left(
        \mathbf{v}_{0,i},a_{t,i}
    \right)
    +
    \mathcal{R}
    \left(
        \Delta\mathbf{v}_{t,i},
        1+\rho(1-a_{t,i})
    \right),
\end{aligned}
\right.
\end{equation}
where $\rho\geq 0$ controls gradient amplification along the adaptation path. CDGR uses the distinct update paths provided by the parallel dual-channel design to support shared learning under agreement and emphasize task-specific adaptation under disagreement. When $a_{t,i}=1$, both paths retain their original gradients. As disagreement increases, CDGR attenuates gradients through the shared projections while placing greater emphasis on the task-specific low-rank adaptation path. Since $\mathcal{R}$ preserves forward values, this coordination leaves the additive attention computation unchanged, and the routing operators can be omitted at inference without additional cost.

\section{Experiments}
Our experiments address the following research questions:
\begin{itemize}[leftmargin=*]
    \item \textbf{RQ1:} How does DuoFormer compare with existing multi-task and scalable recommendation models in predictive performance, parameter count, and computational cost?

    \item \textbf{RQ2:} What predictive benefits and computational costs arise from introducing early task modeling and hub--facet representations into existing baselines?

    \item \textbf{RQ3:} How do DuoFormer's key components contribute to its predictive performance?

    \item \textbf{RQ4:} How does DuoFormer's predictive performance scale with increasing model capacity compared with scalable recommendation baselines?

    \item \textbf{RQ5:} How does DuoFormer affect user engagement and business outcomes in online A/B experiments?
\end{itemize}
\subsection{Experimental Setting}
\subsubsection{Dataset Description}
We evaluate DuoFormer using data from Taobao's search advertising platform, which serves a large user base and records hundreds of millions of clicks per day. The dataset contains approximately 17 billion samples collected over one month. Each sample consists of a user--item pair with multiple feedback labels and associated user and item features. We primarily evaluate performance on the four tasks most critical to our business objectives: overall conversion rate (CVR), direct conversion rate (Dir-CVR), net conversion rate (NCVR), and net direct conversion rate (NDir-CVR). CVR includes direct purchases of the advertised item and indirect purchases of other items from the same store, whereas Dir-CVR considers only direct purchases. NCVR and NDir-CVR are the corresponding refund-adjusted conversion rates.

\subsubsection{Evaluation Protocols}
We report AUC~\cite{fawcett2006roc,zhang2025onetrans} for the four primary conversion tasks and the unweighted mean AUC across all tasks evaluated with this metric.

\subsubsection{Baselines}
We compare DuoFormer with ten baselines, including three conventional multi-task learning models: \textbf{Shared-Bottom}~\cite{caruana1997multitask}, \textbf{MMoE}~\cite{ma2018mmoe}, and \textbf{PLE}~\cite{tang2020ple}; and seven scalable recommendation architectures: \textbf{Wukong}~\cite{zhang2024wukong}, \textbf{RankMixer}~\cite{zhu2025rankmixer}, \textbf{TokenMixer}~\cite{jiang2026tokenmixerlarge}, \textbf{OneTrans}~\cite{zhang2025onetrans}, \textbf{OneRank}~\cite{tang2026onerank}, \textbf{SMES}~\cite{zhang2026smes}, and \textbf{UniFormer}~\cite{chen2025uniformer}.

\subsubsection{Implementation Details}
We train all models from scratch on 32 NVIDIA H20 GPUs with a per-GPU batch size of 4,096. Both sparse and dense parameters are optimized using AdamW~\cite{loshchilov2019decoupled}, with a constant learning rate of $10^{-4}$ and a weight decay of $10^{-6}$. By default, DuoFormer uses $L=3$ interaction layers, token dimension $d=256$, $P=3$ facet tokens per task, adaptation rank $r=8$, and low-rank scaling parameter $\alpha=8$. In CDGR, we set $\tau=0.25$ and $\rho=1$.

\begin{table}[t]
    \centering
    \caption{AUC comparison on the four tasks most critical to our business objectives. Avg. denotes the mean AUC across all tasks evaluated using AUC.}
    \label{tab:overall_performance}
    \footnotesize
    \setlength{\tabcolsep}{2pt}
    \renewcommand{\arraystretch}{1.15}
    \resizebox{\columnwidth}{!}{%
    \begin{tabular}{@{}l*{5}{c}rr@{}}
        \toprule
        Model
        & CVR
        & \shortstack{Dir-\\CVR}
        & NCVR
        & \shortstack{NDir-\\CVR}
        & Avg.
        & \shortstack{\#Params\\(M)}
        & TFLOPs \\
        \midrule
        Shared-Bottom
        & 0.8714 & 0.8803 & 0.8739 & 0.8825 & 0.8670
        & 9.03 & 0.86 \\
        MMoE
        & 0.8791 & 0.8876 & 0.8815 & 0.8896 & 0.8704
        & 17.76 & 0.90 \\
        PLE
        & 0.8789 & 0.8871 & 0.8810 & 0.8893 & 0.8691
        & 17.39 & 0.90 \\
        WuKong
        & 0.8807 & 0.8888 & 0.8827 & 0.8907 & 0.8703
        & 23.63 & 1.16 \\
        RankMixer
        & 0.8816 & 0.8896 & 0.8835 & 0.8915 & 0.8716
        & 40.04 & 1.82 \\
        TokenMixer
        & 0.8829 & 0.8908 & 0.8848 & 0.8927 & 0.8729
        & 51.75 & 1.95 \\
        OneTrans
        & 0.8826 & 0.8906 & 0.8848 & 0.8926 & 0.8738
        & 52.82 & 2.48 \\
        OneRank
        & 0.8827 & 0.8909 & 0.8848 & 0.8929 & 0.8725
        & 53.50 & 10.72 \\
        SMES
        & 0.8831 & 0.8907 & 0.8849 & 0.8930 & 0.8742
        & 59.62 & 1.50 \\
        UniFormer
        & 0.8821 & 0.8902 & 0.8841 & 0.8920 & 0.8718
        & 54.59 & 2.51 \\
        \midrule
        \textbf{DuoFormer}
        & \textbf{0.8839}
        & \textbf{0.8918}
        & \textbf{0.8858}
        & \textbf{0.8936}
        & \textbf{0.8755}
        & 48.84 & 2.42 \\
        \bottomrule
    \end{tabular}%
    }
\end{table}

\subsection{Performance Comparison (RQ1)}

Table~\ref{tab:overall_performance} shows the experimental results, from which we draw four main conclusions.

\textbf{First, DuoFormer achieves the highest AUC on all four reported tasks and the highest average AUC across tasks.} These gains support the effectiveness of our parallel design, which introduces task-dependent feature retrieval and interpretation alongside shared representation learning to enable fine-grained task specialization from early interaction stages.

\textbf{Second, scalable feature modeling provides a stronger foundation for multi-task recommendation.} The gains of scalable baselines highlight the value of expressive shared feature modeling, supporting the allocation of substantial modeling capacity to DuoFormer's shared channel.

\textbf{Third, DuoFormer's advantage over scalable baselines highlights the value of fine-grained task specialization beyond stronger shared encoding.} The gains over SMES and UniFormer support combining early task modeling with task-dependent feature interpretation and multi-view task representations. The improvement over OneRank, which incorporates early task modeling but retains a common feature view and a single task token, further highlights the value of fine-grained task specialization beyond early task modeling.

\textbf{Fourth, DuoFormer enables early task specialization with competitive parameter and computational costs.} DuoFormer achieves the best overall performance with 48.84M dense parameters and 2.42 TFLOPs, at a budget comparable to strong scalable baselines. Compared with OneRank, which also introduces task modeling at early interaction stages, DuoFormer requires 77.4\% fewer FLOPs at a similar parameter scale. This efficiency stems from reusing shared feature computation across tasks and introducing task-dependent feature refinements through lightweight low-rank adaptation.

\begin{figure}[t]
\centering
\includegraphics[width=1.0\linewidth]{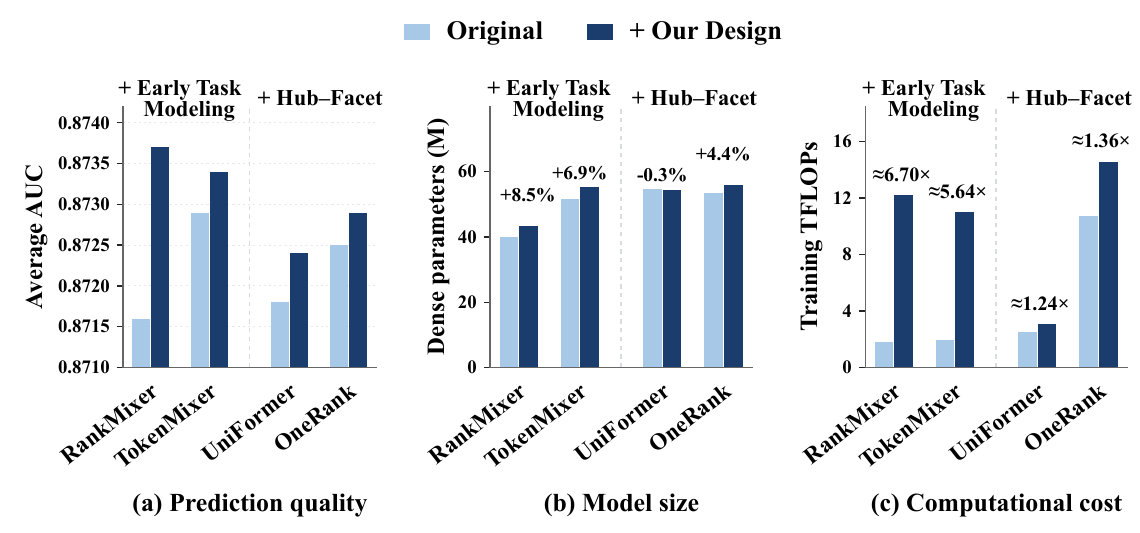}
\caption{Effects of early task modeling on RankMixer and TokenMixer and hub--facet modeling on UniFormer and OneRank, measured by (a) average AUC, (b) dense parameter count, and (c) training FLOPs.}
\label{fig:design_validation}
\end{figure}

\subsection{Validation of Design Motivations (RQ2)}
\label{sec:design_validation}

We examine DuoFormer's core motivations by introducing early task modeling and hub--facet representations into existing baselines, targeting delayed task specialization and coarse-grained task representations. These modifications allow us to assess the predictive benefits and computational overhead of addressing these limitations in existing models, as summarized in Figure~\ref{fig:design_validation}.

\textbf{Effectiveness and Cost of Early Task Modeling.}
To validate our motivation for early task specialization, we introduce task-specific modeling from the first mixer layer in RankMixer and TokenMixer. We aim to keep parameter counts close to those of the original models to reduce the influence of model capacity on AUC comparisons. Both variants achieve higher average AUC, supporting the effectiveness of developing task-dependent feature representations during early feature interaction. However, executing the mixer separately for each task duplicates feature-interaction computation and substantially increases training FLOPs. This trade-off motivates DuoFormer's integration of lightweight task-specific modeling into a reusable shared backbone.

\textbf{Effectiveness of Hub--Facet Task Modeling.}
To validate our motivation for multi-view task modeling, we expand each task's representation in UniFormer and OneRank into a hub and multiple facet tokens. The facets capture distinct task-relevant information, while the hub coordinates and integrates their representations. Both variants improve average AUC with comparable dense parameter counts, supporting the value of separating multi-view information extraction from task-level integration.

\begin{figure}[t]
\centering
\includegraphics[width=1.0\linewidth]{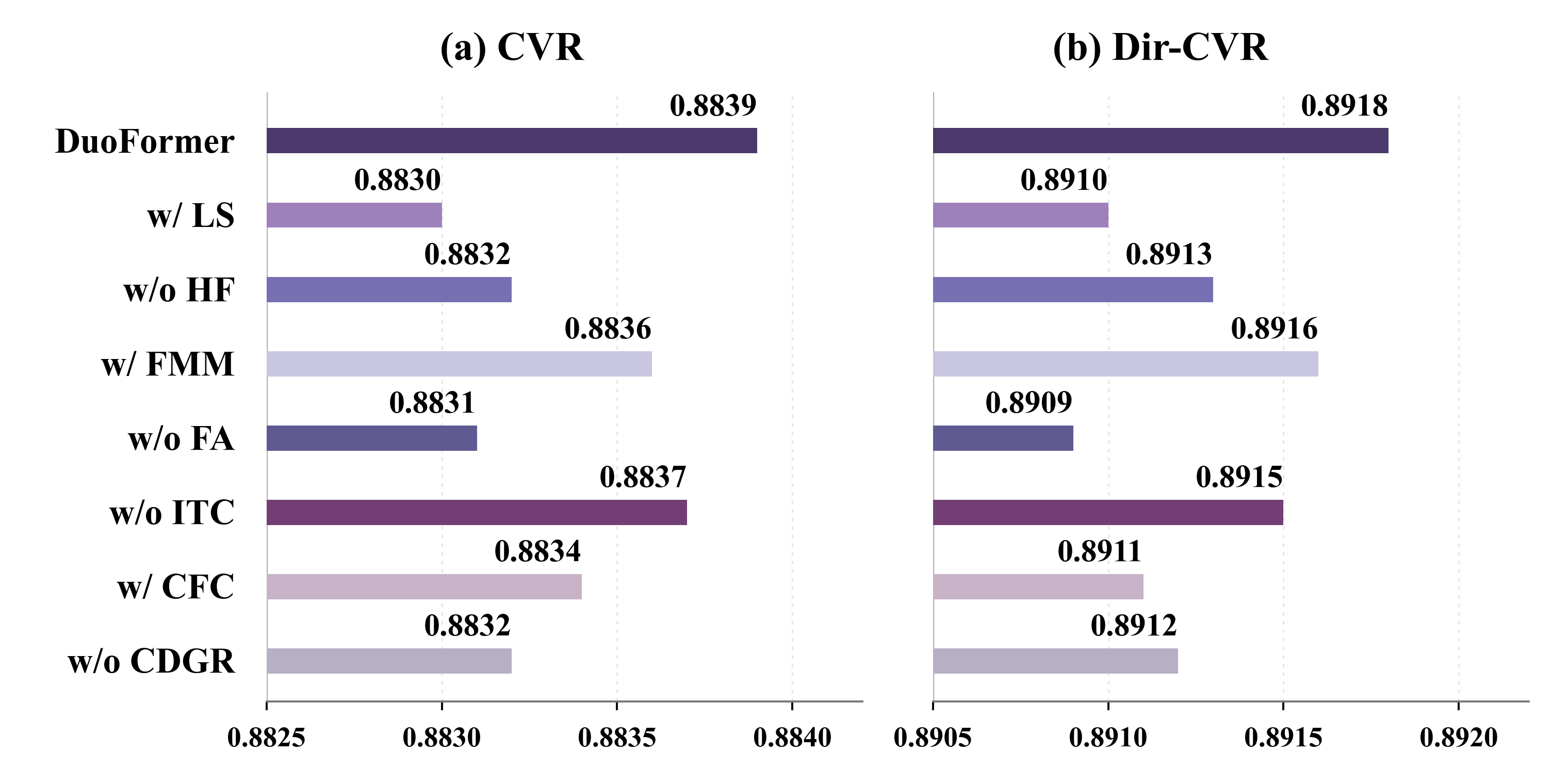}
\caption{Ablation results on CVR and Dir-CVR.}
\label{fig:ablation}
\end{figure}

\subsection{Ablation Study (RQ3)}
\label{sec:ablation}

We evaluate seven variants to examine the contributions of our key components. \textbf{w/ LS} applies \textbf{L}ate \textbf{S}pecialization by introducing task cross-attention only after shared feature encoding is complete. \textbf{w/o HF} replaces the \textbf{H}ub--\textbf{F}acet hierarchy with a single token per task. \textbf{w/ FMM} adopts \textbf{F}lat \textbf{M}ulti-Token \textbf{M}odeling with the same token count but no distinct hub--facet roles, using attention pooling for task-level communication and prediction. \textbf{w/o FA} removes \textbf{F}eature \textbf{A}daptation by omitting task-specific key and value increments while retaining task-dependent queries. \textbf{w/o ITC} disables \textbf{I}nter-\textbf{T}ask \textbf{C}ommunication between hubs. \textbf{w/ CFC} adds \textbf{C}ross-Task \textbf{F}acet \textbf{C}ommunication by allowing attention between facets from different tasks. \textbf{w/o CDGR} removes \textbf{C}onsensus--\textbf{D}isagreement \textbf{G}radient \textbf{R}outing and uses standard backpropagation. Figure~\ref{fig:ablation} shows that all seven variants underperform DuoFormer on both tasks, with the following implications for our design choices.

\textbf{Early Task Specialization and Feature Adaptation.} w/ LS and w/o FA yield the largest performance drops across the two tasks. The degradation of w/ LS supports developing task-specific representations alongside shared feature learning from early interaction stages. Despite retaining task-dependent facet queries, w/o FA performs worse, highlighting the value of adapting both feature matching and content representations to task-specific requirements.

\textbf{Hierarchical Multi-View Modeling.} w/ FMM outperforms w/o HF, supporting the preservation of diverse task factors through multiple modeling views. DuoFormer further outperforms w/ FMM with the same number of tokens per task, highlighting the value of hub--facet roles in separating fine-grained information extraction from task-level coordination.

\textbf{Selective Cross-Task Communication.} The degradation of w/o ITC supports exchanging complementary task information through hubs. w/ CFC also underperforms DuoFormer, suggesting that direct communication between facets from different tasks can propagate task-private information and blur task-specific distinctions. These results support exchanging consolidated task information through hubs while keeping fine-grained facet representations local to each task.

\textbf{Optimization Aligned with the Parallel Design.} w/o CDGR lowers AUC on both tasks despite retaining the same forward architecture, demonstrating the benefit of coordinating shared and task-specific updates. These gains support preserving compatible shared updates and using gradient disagreement to strengthen task-specific adaptation, improving prediction without additional inference computation.

\begin{figure}[t]
\centering
\includegraphics[width=1.0\linewidth]{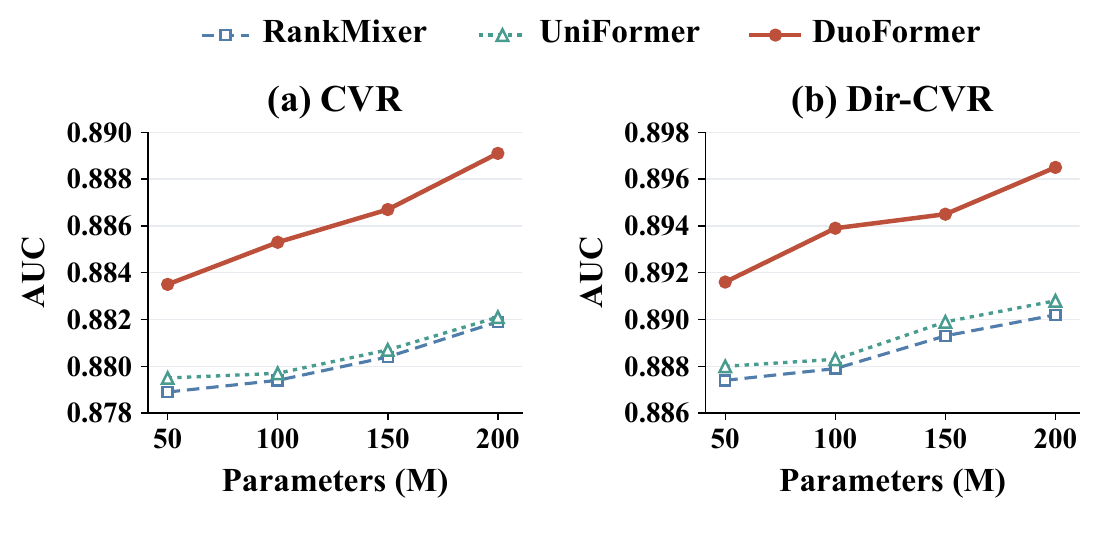}
\caption{Scaling performance of DuoFormer, RankMixer, and UniFormer on the CVR and Dir-CVR tasks.}
\label{fig:scaling_analysis}
\end{figure}

\subsection{Scaling Analysis (RQ4)}

Figure~\ref{fig:scaling_analysis} shows that DuoFormer consistently improves as model capacity increases and outperforms RankMixer and UniFormer at every evaluated scale on both tasks. DuoFormer also exhibits steeper scaling curves than the baselines, achieving larger performance gains as the parameter count increases over the evaluated range. These results support our motivation to combine scalable shared feature learning with fine-grained task specialization. Their parallel integration enables DuoFormer to translate increased capacity into stronger recommendation performance while maintaining task-specific feature retrieval and interpretation.

\subsection{Online A/B Testing (RQ5)}
We deployed DuoFormer in Taobao's sponsored search ranking system for a 7-day online A/B test using 5\% of traffic. DuoFormer increased conversion rate (CVR) by 1.40\% and gross merchandise value (GMV) by 1.41\%. The add-to-favorites and add-to-cart rates also improved by 2.62\% and 1.87\%, respectively. These results demonstrate the practical benefits of DuoFormer in improving user engagement and business outcomes on large-scale e-commerce platforms.

\section{Conclusion}

We presented DuoFormer, a scalable architecture for multi-task recommendation that develops fine-grained task representations alongside shared feature learning from the earliest interaction stages. Experiments on large-scale industrial data demonstrate consistent predictive gains, competitive computational costs, and favorable scaling behavior.

\bibliographystyle{ACM-Reference-Format}
\bibliography{references}


\begin{thebibliography}{30}


\ifx \showCODEN    \undefined \def \showCODEN     #1{\unskip}     \fi
\ifx \showISBNx    \undefined \def \showISBNx     #1{\unskip}     \fi
\ifx \showISBNxiii \undefined \def \showISBNxiii  #1{\unskip}     \fi
\ifx \showISSN     \undefined \def \showISSN      #1{\unskip}     \fi
\ifx \showLCCN     \undefined \def \showLCCN      #1{\unskip}     \fi
\ifx \shownote     \undefined \def \shownote      #1{#1}          \fi
\ifx \showarticletitle \undefined \def \showarticletitle #1{#1}   \fi
\ifx \showURL      \undefined \def \showURL       {\relax}        \fi
\providecommand\bibfield[2]{#2}
\providecommand\bibinfo[2]{#2}
\providecommand\natexlab[1]{#1}
\providecommand\showeprint[2][]{arXiv:#2}

\bibitem[Caruana(1997)]%
        {caruana1997multitask}
\bibfield{author}{\bibinfo{person}{Rich Caruana}.}
  \bibinfo{year}{1997}\natexlab{}.
\newblock \showarticletitle{Multitask Learning}.
\newblock \bibinfo{journal}{\emph{Machine Learning}}  \bibinfo{volume}{28}
  (\bibinfo{year}{1997}), \bibinfo{pages}{41--75}.
\newblock
\href{https://doi.org/10.1023/A:1007379606734}{doi:\nolinkurl{10.1023/A:1007379606734}}


\bibitem[Chen et~al\mbox{.}(2026)]%
        {chen2025uniformer}
\bibfield{author}{\bibinfo{person}{Bo Chen}, \bibinfo{person}{Jinlong Jiao},
  \bibinfo{person}{Tijian Hu}, \bibinfo{person}{Ruihao Zhang},
  \bibinfo{person}{Yanzhi Liu}, \bibinfo{person}{Chenghou Jin},
  \bibinfo{person}{Qinglin Jia}, \bibinfo{person}{Baixuan He},
  \bibinfo{person}{Hechang Pan}, \bibinfo{person}{Yiwu Liu},
  \bibinfo{person}{Jian Liang}, \bibinfo{person}{Chaoyi Ma},
  \bibinfo{person}{Ruiming Tang}, \bibinfo{person}{Han Li}, {and}
  \bibinfo{person}{Kun Gai}.} \bibinfo{year}{2026}\natexlab{}.
\newblock \showarticletitle{{UniFormer}: Efficient and Unified Model-Centric
  Scaling for Industrial Recommendation}.
\newblock \bibinfo{journal}{\emph{arXiv preprint arXiv:2606.27058}}
  (\bibinfo{year}{2026}).
\newblock
\urldef\tempurl%
\url{https://arxiv.org/abs/2606.27058}
\showURL{%
\tempurl}


\bibitem[Chen et~al\mbox{.}(2025)]%
        {chen2025mal}
\bibfield{author}{\bibinfo{person}{Sishuo Chen}, \bibinfo{person}{Zhangming
  Chan}, \bibinfo{person}{Xiang-Rong Sheng}, \bibinfo{person}{Lei Zhang},
  \bibinfo{person}{Sheng Chen}, \bibinfo{person}{Chenghuan Hou},
  \bibinfo{person}{Han Zhu}, \bibinfo{person}{Jian Xu}, {and}
  \bibinfo{person}{Bo Zheng}.} \bibinfo{year}{2025}\natexlab{}.
\newblock \bibinfo{title}{See Beyond a Single View: Multi-Attribution Learning
  Leads to Better Conversion Rate Prediction}.
\newblock
\showeprint[arxiv]{2508.15217}~[cs.LG]
\urldef\tempurl%
\url{https://arxiv.org/abs/2508.15217}
\showURL{%
\tempurl}


\bibitem[Covington et~al\mbox{.}(2016)]%
        {covington2016youtube}
\bibfield{author}{\bibinfo{person}{Paul Covington}, \bibinfo{person}{Jay
  Adams}, {and} \bibinfo{person}{Emre Sargin}.}
  \bibinfo{year}{2016}\natexlab{}.
\newblock \showarticletitle{Deep Neural Networks for {YouTube}
  Recommendations}. In \bibinfo{booktitle}{\emph{Proceedings of the 10th ACM
  Conference on Recommender Systems}}. \bibinfo{pages}{191--198}.
\newblock
\href{https://doi.org/10.1145/2959100.2959190}{doi:\nolinkurl{10.1145/2959100.2959190}}


\bibitem[Fawcett(2006)]%
        {fawcett2006roc}
\bibfield{author}{\bibinfo{person}{Tom Fawcett}.}
  \bibinfo{year}{2006}\natexlab{}.
\newblock \showarticletitle{An Introduction to {ROC} Analysis}.
\newblock \bibinfo{journal}{\emph{Pattern Recognition Letters}}
  \bibinfo{volume}{27}, \bibinfo{number}{8} (\bibinfo{year}{2006}),
  \bibinfo{pages}{861--874}.
\newblock
\href{https://doi.org/10.1016/j.patrec.2005.10.010}{doi:\nolinkurl{10.1016/j.patrec.2005.10.010}}


\bibitem[Fernando et~al\mbox{.}(2023)]%
        {fernando2023mitigating}
\bibfield{author}{\bibinfo{person}{Heshan~Devaka Fernando},
  \bibinfo{person}{Han Shen}, \bibinfo{person}{Miao Liu},
  \bibinfo{person}{Subhajit Chaudhury}, \bibinfo{person}{Keerthiram Murugesan},
  {and} \bibinfo{person}{Tianyi Chen}.} \bibinfo{year}{2023}\natexlab{}.
\newblock \showarticletitle{Mitigating Gradient Bias in Multi-objective
  Learning: A Provably Convergent Approach}. In \bibinfo{booktitle}{\emph{The
  Eleventh International Conference on Learning Representations}}.
\newblock
\urldef\tempurl%
\url{https://openreview.net/forum?id=dLAYGdKTi2}
\showURL{%
\tempurl}


\bibitem[Gao et~al\mbox{.}(2022)]%
        {gao2022kuairand}
\bibfield{author}{\bibinfo{person}{Chongming Gao}, \bibinfo{person}{Shijun Li},
  \bibinfo{person}{Yuan Zhang}, \bibinfo{person}{Jiawei Chen},
  \bibinfo{person}{Biao Li}, \bibinfo{person}{Wenqiang Lei},
  \bibinfo{person}{Peng Jiang}, {and} \bibinfo{person}{Xiangnan He}.}
  \bibinfo{year}{2022}\natexlab{}.
\newblock \showarticletitle{{KuaiRand}: An Unbiased Sequential Recommendation
  Dataset with Randomly Exposed Videos}. In
  \bibinfo{booktitle}{\emph{Proceedings of the 31st ACM International
  Conference on Information \& Knowledge Management}}.
  \bibinfo{pages}{3953--3957}.
\newblock
\href{https://doi.org/10.1145/3511808.3557624}{doi:\nolinkurl{10.1145/3511808.3557624}}


\bibitem[Ha et~al\mbox{.}(2026)]%
        {ha2026unimixer}
\bibfield{author}{\bibinfo{person}{Mingming Ha}, \bibinfo{person}{Guanchen
  Wang}, \bibinfo{person}{Linxun Chen}, \bibinfo{person}{Xuan Rao},
  \bibinfo{person}{Yuexin Shi}, \bibinfo{person}{Tianbao Ma},
  \bibinfo{person}{Zhaojie Liu}, \bibinfo{person}{Yunqian Fan},
  \bibinfo{person}{Zilong Lu}, \bibinfo{person}{Yanan Niu},
  \bibinfo{person}{Han Li}, {and} \bibinfo{person}{Kun Gai}.}
  \bibinfo{year}{2026}\natexlab{}.
\newblock \bibinfo{title}{UniMixer: A Unified Architecture for Scaling Laws in
  Recommendation Systems}.
\newblock
\showeprint[arxiv]{2604.00590}~[cs.IR]
\urldef\tempurl%
\url{https://arxiv.org/abs/2604.00590}
\showURL{%
\tempurl}


\bibitem[Hoffmann et~al\mbox{.}(2022)]%
        {hoffmann2022training}
\bibfield{author}{\bibinfo{person}{Jordan Hoffmann}, \bibinfo{person}{Sebastian
  Borgeaud}, \bibinfo{person}{Arthur Mensch}, \bibinfo{person}{Elena
  Buchatskaya}, \bibinfo{person}{Trevor Cai}, \bibinfo{person}{Eliza
  Rutherford}, \bibinfo{person}{Diego de Las~Casas}, \bibinfo{person}{Lisa~Anne
  Hendricks}, \bibinfo{person}{Johannes Welbl}, \bibinfo{person}{Aidan Clark},
  \bibinfo{person}{Tom Hennigan}, \bibinfo{person}{Eric Noland},
  \bibinfo{person}{Katie Millican}, \bibinfo{person}{George van~den Driessche},
  \bibinfo{person}{Bogdan Damoc}, \bibinfo{person}{Aurelia Guy},
  \bibinfo{person}{Simon Osindero}, \bibinfo{person}{Karen Simonyan},
  \bibinfo{person}{Erich Elsen}, \bibinfo{person}{Jack~W. Rae},
  \bibinfo{person}{Oriol Vinyals}, {and} \bibinfo{person}{Laurent Sifre}.}
  \bibinfo{year}{2022}\natexlab{}.
\newblock \showarticletitle{Training Compute-Optimal Large Language Models}. In
  \bibinfo{booktitle}{\emph{Advances in Neural Information Processing
  Systems}}, Vol.~\bibinfo{volume}{35}. \bibinfo{pages}{30016--30030}.
\newblock


\bibitem[Huang et~al\mbox{.}(2026)]%
        {huang2026mixformer}
\bibfield{author}{\bibinfo{person}{Xu Huang}, \bibinfo{person}{Hao Zhang},
  \bibinfo{person}{Zhifang Fan}, \bibinfo{person}{Yunwen Huang},
  \bibinfo{person}{Zhuoxing Wei}, \bibinfo{person}{Zheng Chai},
  \bibinfo{person}{Jinan Ni}, \bibinfo{person}{Yuchao Zheng}, {and}
  \bibinfo{person}{Qiwei Chen}.} \bibinfo{year}{2026}\natexlab{}.
\newblock \showarticletitle{{MixFormer}: Co-Scaling Up Dense and Sequence in
  Industrial Recommenders}. In \bibinfo{booktitle}{\emph{Proceedings of the
  32nd ACM SIGKDD Conference on Knowledge Discovery and Data Mining}}.
  \bibinfo{pages}{7422--7431}.
\newblock
\href{https://doi.org/10.1145/3770855.3818447}{doi:\nolinkurl{10.1145/3770855.3818447}}


\bibitem[Jiang et~al\mbox{.}(2026)]%
        {jiang2026tokenmixerlarge}
\bibfield{author}{\bibinfo{person}{Yuchen Jiang}, \bibinfo{person}{Jie Zhu},
  \bibinfo{person}{Xintian Han}, \bibinfo{person}{Hui Lu},
  \bibinfo{person}{Kunmin Bai}, \bibinfo{person}{Mingyu Yang},
  \bibinfo{person}{Shikang Wu}, \bibinfo{person}{Ruihao Zhang},
  \bibinfo{person}{Wenlin Zhao}, \bibinfo{person}{Shipeng Bai},
  \bibinfo{person}{Sijin Zhou}, \bibinfo{person}{Huizhi Yang},
  \bibinfo{person}{Tianyi Liu}, \bibinfo{person}{Wenda Liu},
  \bibinfo{person}{Ziyan Gong}, \bibinfo{person}{Haoran Ding},
  \bibinfo{person}{Zheng Chai}, \bibinfo{person}{Deping Xie},
  \bibinfo{person}{Zhe Chen}, \bibinfo{person}{Yuchao Zheng}, {and}
  \bibinfo{person}{Peng Xu}.} \bibinfo{year}{2026}\natexlab{}.
\newblock \showarticletitle{{TokenMixer-Large}: Scaling Up Large Ranking Models
  in Industrial Recommenders}. In \bibinfo{booktitle}{\emph{Proceedings of the
  32nd ACM SIGKDD Conference on Knowledge Discovery and Data Mining}}.
  \bibinfo{pages}{7489--7500}.
\newblock
\href{https://doi.org/10.1145/3770855.3818494}{doi:\nolinkurl{10.1145/3770855.3818494}}


\bibitem[Kaplan et~al\mbox{.}(2020)]%
        {kaplan2020scaling}
\bibfield{author}{\bibinfo{person}{Jared Kaplan}, \bibinfo{person}{Sam
  McCandlish}, \bibinfo{person}{Tom Henighan}, \bibinfo{person}{Tom~B. Brown},
  \bibinfo{person}{Benjamin Chess}, \bibinfo{person}{Rewon Child},
  \bibinfo{person}{Scott Gray}, \bibinfo{person}{Alec Radford},
  \bibinfo{person}{Jeffrey Wu}, {and} \bibinfo{person}{Dario Amodei}.}
  \bibinfo{year}{2020}\natexlab{}.
\newblock \showarticletitle{Scaling Laws for Neural Language Models}.
\newblock \bibinfo{journal}{\emph{arXiv preprint arXiv:2001.08361}}
  (\bibinfo{year}{2020}).
\newblock
\urldef\tempurl%
\url{https://arxiv.org/abs/2001.08361}
\showURL{%
\tempurl}


\bibitem[Li et~al\mbox{.}(2023)]%
        {li2023adatt}
\bibfield{author}{\bibinfo{person}{Danwei Li}, \bibinfo{person}{Zhengyu Zhang},
  \bibinfo{person}{Siyang Yuan}, \bibinfo{person}{Mingze Gao},
  \bibinfo{person}{Weilin Zhang}, \bibinfo{person}{Chaofei Yang},
  \bibinfo{person}{Xi Liu}, {and} \bibinfo{person}{Jiyan Yang}.}
  \bibinfo{year}{2023}\natexlab{}.
\newblock \showarticletitle{{AdaTT}: Adaptive Task-to-Task Fusion Network for
  Multitask Learning in Recommendations}.
\newblock \bibinfo{journal}{\emph{arXiv preprint arXiv:2304.04959}}
  (\bibinfo{year}{2023}).
\newblock
\urldef\tempurl%
\url{https://arxiv.org/abs/2304.04959}
\showURL{%
\tempurl}


\bibitem[Li et~al\mbox{.}(2026)]%
        {li2026spectral}
\bibfield{author}{\bibinfo{person}{Xiangming Li}, \bibinfo{person}{Hua Chu},
  \bibinfo{person}{Chengyu Feng}, \bibinfo{person}{Jianan Li},
  \bibinfo{person}{Yangtao Zhou}, \bibinfo{person}{Qingshan Li}, {and}
  \bibinfo{person}{Wanqiang Yang}.} \bibinfo{year}{2026}\natexlab{}.
\newblock \showarticletitle{A Spectral Heterogeneous Diffusion Framework for
  Knowledge-aware Recommendation}. In \bibinfo{booktitle}{\emph{Proceedings of
  the Nineteenth ACM International Conference on Web Search and Data Mining}}.
  \bibinfo{pages}{366--376}.
\newblock


\bibitem[Liu et~al\mbox{.}(2021)]%
        {liu2021conflict}
\bibfield{author}{\bibinfo{person}{Bo Liu}, \bibinfo{person}{Xingchao Liu},
  \bibinfo{person}{Xiaojie Jin}, \bibinfo{person}{Peter Stone}, {and}
  \bibinfo{person}{Qiang Liu}.} \bibinfo{year}{2021}\natexlab{}.
\newblock \showarticletitle{Conflict-Averse Gradient Descent for Multi-task
  Learning}. In \bibinfo{booktitle}{\emph{Advances in Neural Information
  Processing Systems}}, Vol.~\bibinfo{volume}{34}. \bibinfo{publisher}{Curran
  Associates, Inc.}, \bibinfo{pages}{18878--18890}.
\newblock
\urldef\tempurl%
\url{https://proceedings.neurips.cc/paper_files/paper/2021/file/9d27fdf2477ffbff837d73ef7ae23db9-Paper.pdf}
\showURL{%
\tempurl}


\bibitem[Loshchilov and Hutter(2019)]%
        {loshchilov2019decoupled}
\bibfield{author}{\bibinfo{person}{Ilya Loshchilov} {and}
  \bibinfo{person}{Frank Hutter}.} \bibinfo{year}{2019}\natexlab{}.
\newblock \showarticletitle{Decoupled Weight Decay Regularization}. In
  \bibinfo{booktitle}{\emph{International Conference on Learning
  Representations}}.
\newblock
\urldef\tempurl%
\url{https://openreview.net/forum?id=Bkg6RiCqY7}
\showURL{%
\tempurl}


\bibitem[Ma et~al\mbox{.}(2018b)]%
        {ma2018mmoe}
\bibfield{author}{\bibinfo{person}{Jiaqi Ma}, \bibinfo{person}{Zhe Zhao},
  \bibinfo{person}{Xinyang Yi}, \bibinfo{person}{Jilin Chen},
  \bibinfo{person}{Lichan Hong}, {and} \bibinfo{person}{Ed~H. Chi}.}
  \bibinfo{year}{2018}\natexlab{b}.
\newblock \showarticletitle{Modeling Task Relationships in Multi-task Learning
  with Multi-gate Mixture-of-Experts}. In \bibinfo{booktitle}{\emph{Proceedings
  of the 24th ACM SIGKDD International Conference on Knowledge Discovery and
  Data Mining}}. \bibinfo{pages}{1930--1939}.
\newblock
\href{https://doi.org/10.1145/3219819.3220007}{doi:\nolinkurl{10.1145/3219819.3220007}}


\bibitem[Ma et~al\mbox{.}(2018a)]%
        {ma2018esmm}
\bibfield{author}{\bibinfo{person}{Xiao Ma}, \bibinfo{person}{Liqin Zhao},
  \bibinfo{person}{Guan Huang}, \bibinfo{person}{Zhi Wang},
  \bibinfo{person}{Zelin Hu}, \bibinfo{person}{Xiaoqiang Zhu}, {and}
  \bibinfo{person}{Kun Gai}.} \bibinfo{year}{2018}\natexlab{a}.
\newblock \showarticletitle{Entire Space Multi-Task Model: An Effective
  Approach for Estimating Post-Click Conversion Rate}. In
  \bibinfo{booktitle}{\emph{Proceedings of the 41st International ACM SIGIR
  Conference on Research \& Development in Information Retrieval}}.
  \bibinfo{pages}{1137--1140}.
\newblock
\href{https://doi.org/10.1145/3209978.3210104}{doi:\nolinkurl{10.1145/3209978.3210104}}


\bibitem[Su et~al\mbox{.}(2024)]%
        {su2024stem}
\bibfield{author}{\bibinfo{person}{Liangcai Su}, \bibinfo{person}{Junwei Pan},
  \bibinfo{person}{Ximei Wang}, \bibinfo{person}{Xi Xiao},
  \bibinfo{person}{Shijie Quan}, \bibinfo{person}{Xihua Chen}, {and}
  \bibinfo{person}{Jie Jiang}.} \bibinfo{year}{2024}\natexlab{}.
\newblock \showarticletitle{{STEM}: Unleashing the Power of Embeddings for
  Multi-Task Recommendation}.
\newblock \bibinfo{journal}{\emph{Proceedings of the AAAI Conference on
  Artificial Intelligence}} \bibinfo{volume}{38}, \bibinfo{number}{8}
  (\bibinfo{year}{2024}), \bibinfo{pages}{9002--9010}.
\newblock
\href{https://doi.org/10.1609/aaai.v38i8.28749}{doi:\nolinkurl{10.1609/aaai.v38i8.28749}}


\bibitem[Tang et~al\mbox{.}(2020)]%
        {tang2020ple}
\bibfield{author}{\bibinfo{person}{Hongyan Tang}, \bibinfo{person}{Junning
  Liu}, \bibinfo{person}{Ming Zhao}, {and} \bibinfo{person}{Xudong Gong}.}
  \bibinfo{year}{2020}\natexlab{}.
\newblock \showarticletitle{Progressive Layered Extraction ({PLE}): A Novel
  Multi-Task Learning ({MTL}) Model for Personalized Recommendations}. In
  \bibinfo{booktitle}{\emph{Proceedings of the 14th ACM Conference on
  Recommender Systems}}. \bibinfo{pages}{269--278}.
\newblock
\href{https://doi.org/10.1145/3383313.3412236}{doi:\nolinkurl{10.1145/3383313.3412236}}


\bibitem[Tang et~al\mbox{.}(2026)]%
        {tang2026onerank}
\bibfield{author}{\bibinfo{person}{Jiakai Tang}, \bibinfo{person}{Sunhao Dai},
  \bibinfo{person}{Kun Wang}, \bibinfo{person}{Zhiluohan Guo},
  \bibinfo{person}{Yu Zhao}, \bibinfo{person}{Cong Fu}, \bibinfo{person}{Kangle
  Wu}, \bibinfo{person}{Yabo Ni}, \bibinfo{person}{Anxiang Zeng},
  \bibinfo{person}{Xu Chen}, {and} \bibinfo{person}{Jun Xu}.}
  \bibinfo{year}{2026}\natexlab{}.
\newblock \showarticletitle{{OneRank}: Unified Transformer-Native Ranking
  Architecture for Multi-Task Recommendation}. In
  \bibinfo{booktitle}{\emph{Proceedings of the 32nd ACM SIGKDD Conference on
  Knowledge Discovery and Data Mining}}. \bibinfo{pages}{8053--8064}.
\newblock
\href{https://doi.org/10.1145/3770855.3818457}{doi:\nolinkurl{10.1145/3770855.3818457}}


\bibitem[Vaswani et~al\mbox{.}(2017)]%
        {vaswani2017attention}
\bibfield{author}{\bibinfo{person}{Ashish Vaswani}, \bibinfo{person}{Noam
  Shazeer}, \bibinfo{person}{Niki Parmar}, \bibinfo{person}{Jakob Uszkoreit},
  \bibinfo{person}{Llion Jones}, \bibinfo{person}{Aidan~N. Gomez},
  \bibinfo{person}{Lukasz Kaiser}, {and} \bibinfo{person}{Illia Polosukhin}.}
  \bibinfo{year}{2017}\natexlab{}.
\newblock \showarticletitle{Attention Is All You Need}. In
  \bibinfo{booktitle}{\emph{Advances in Neural Information Processing
  Systems}}, Vol.~\bibinfo{volume}{30}.
\newblock


\bibitem[Yu et~al\mbox{.}(2020)]%
        {yu2020gradient}
\bibfield{author}{\bibinfo{person}{Tianhe Yu}, \bibinfo{person}{Saurabh Kumar},
  \bibinfo{person}{Abhishek Gupta}, \bibinfo{person}{Sergey Levine},
  \bibinfo{person}{Karol Hausman}, {and} \bibinfo{person}{Chelsea Finn}.}
  \bibinfo{year}{2020}\natexlab{}.
\newblock \showarticletitle{Gradient Surgery for Multi-Task Learning}. In
  \bibinfo{booktitle}{\emph{Advances in Neural Information Processing
  Systems}}, Vol.~\bibinfo{volume}{33}. \bibinfo{publisher}{Curran Associates,
  Inc.}, \bibinfo{pages}{5824--5836}.
\newblock
\urldef\tempurl%
\url{https://proceedings.neurips.cc/paper_files/paper/2020/file/3fe78a8acf5fda99de95303940a2420c-Paper.pdf}
\showURL{%
\tempurl}


\bibitem[Zhai et~al\mbox{.}(2024)]%
        {zhai2024actions}
\bibfield{author}{\bibinfo{person}{Jiaqi Zhai}, \bibinfo{person}{Lucy Liao},
  \bibinfo{person}{Xing Liu}, \bibinfo{person}{Yueming Wang},
  \bibinfo{person}{Rui Li}, \bibinfo{person}{Xuan Cao}, \bibinfo{person}{Leon
  Gao}, \bibinfo{person}{Zhaojie Gong}, \bibinfo{person}{Fangda Gu},
  \bibinfo{person}{Jiayuan He}, \bibinfo{person}{Yinghai Lu}, {and}
  \bibinfo{person}{Yu Shi}.} \bibinfo{year}{2024}\natexlab{}.
\newblock \showarticletitle{Actions Speak Louder than Words: Trillion-Parameter
  Sequential Transducers for Generative Recommendations}. In
  \bibinfo{booktitle}{\emph{Proceedings of the 41st International Conference on
  Machine Learning}} \emph{(\bibinfo{series}{Proceedings of Machine Learning
  Research}, Vol.~\bibinfo{volume}{235})}. \bibinfo{publisher}{PMLR},
  \bibinfo{pages}{58484--58509}.
\newblock
\urldef\tempurl%
\url{https://proceedings.mlr.press/v235/zhai24a.html}
\showURL{%
\tempurl}


\bibitem[Zhang et~al\mbox{.}(2024)]%
        {zhang2024wukong}
\bibfield{author}{\bibinfo{person}{Buyun Zhang}, \bibinfo{person}{Liang Luo},
  \bibinfo{person}{Yuxin Chen}, \bibinfo{person}{Jade Nie}, \bibinfo{person}{Xi
  Liu}, \bibinfo{person}{Daifeng Guo}, \bibinfo{person}{Yanli Zhao},
  \bibinfo{person}{Shen Li}, \bibinfo{person}{Yuchen Hao},
  \bibinfo{person}{Yantao Yao}, \bibinfo{person}{Guna Lakshminarayanan},
  \bibinfo{person}{Ellie~Dingqiao Wen}, \bibinfo{person}{Jongsoo Park},
  \bibinfo{person}{Maxim Naumov}, {and} \bibinfo{person}{Wenlin Chen}.}
  \bibinfo{year}{2024}\natexlab{}.
\newblock \showarticletitle{Wukong: Towards a Scaling Law for Large-Scale
  Recommendation}. In \bibinfo{booktitle}{\emph{Proceedings of the 41st
  International Conference on Machine Learning}}
  \emph{(\bibinfo{series}{Proceedings of Machine Learning Research},
  Vol.~\bibinfo{volume}{235})}. \bibinfo{pages}{59421--59434}.
\newblock
\urldef\tempurl%
\url{https://arxiv.org/abs/2403.02545}
\showURL{%
\tempurl}


\bibitem[Zhang et~al\mbox{.}(2026a)]%
        {zhang2026smes}
\bibfield{author}{\bibinfo{person}{Yukun Zhang}, \bibinfo{person}{Si Dong},
  \bibinfo{person}{Xu Wang}, \bibinfo{person}{Bo Chen},
  \bibinfo{person}{Qinglin Jia}, \bibinfo{person}{Shengzhe Wang},
  \bibinfo{person}{Jinlong Jiao}, \bibinfo{person}{Runhan Li},
  \bibinfo{person}{Jiaqiang Liu}, \bibinfo{person}{Chaoyi Ma},
  \bibinfo{person}{Ruiming Tang}, {and} \bibinfo{person}{Guorui Zhou}.}
  \bibinfo{year}{2026}\natexlab{a}.
\newblock \showarticletitle{{SMES}: Towards Scalable Multi-Task Recommendation
  via Expert Sparsity}. In \bibinfo{booktitle}{\emph{Proceedings of the 32nd
  ACM SIGKDD Conference on Knowledge Discovery and Data Mining}}.
  \bibinfo{pages}{8522--8531}.
\newblock
\href{https://doi.org/10.1145/3770855.3818483}{doi:\nolinkurl{10.1145/3770855.3818483}}


\bibitem[Zhang et~al\mbox{.}(2026b)]%
        {zhang2025onetrans}
\bibfield{author}{\bibinfo{person}{Zhaoqi Zhang}, \bibinfo{person}{Haolei Pei},
  \bibinfo{person}{Jun Guo}, \bibinfo{person}{Tianyu Wang},
  \bibinfo{person}{Yufei Feng}, \bibinfo{person}{Hui Sun},
  \bibinfo{person}{Shaowei Liu}, {and} \bibinfo{person}{Aixin Sun}.}
  \bibinfo{year}{2026}\natexlab{b}.
\newblock \showarticletitle{{OneTrans}: Unified Feature Interaction and
  Sequence Modeling with One Transformer in Industrial Recommender}. In
  \bibinfo{booktitle}{\emph{Proceedings of the ACM Web Conference 2026}}.
\newblock
\showeprint[arxiv]{2510.26104}
\urldef\tempurl%
\url{https://arxiv.org/abs/2510.26104}
\showURL{%
\tempurl}


\bibitem[Zhou et~al\mbox{.}(2018)]%
        {zhou2018din}
\bibfield{author}{\bibinfo{person}{Guorui Zhou}, \bibinfo{person}{Xiaoqiang
  Zhu}, \bibinfo{person}{Chenru Song}, \bibinfo{person}{Ying Fan},
  \bibinfo{person}{Han Zhu}, \bibinfo{person}{Xiao Ma},
  \bibinfo{person}{Yanghui Yan}, \bibinfo{person}{Junqi Jin},
  \bibinfo{person}{Han Li}, {and} \bibinfo{person}{Kun Gai}.}
  \bibinfo{year}{2018}\natexlab{}.
\newblock \showarticletitle{Deep Interest Network for Click-Through Rate
  Prediction}. In \bibinfo{booktitle}{\emph{Proceedings of the 24th ACM SIGKDD
  International Conference on Knowledge Discovery and Data Mining}}.
  \bibinfo{pages}{1059--1068}.
\newblock
\href{https://doi.org/10.1145/3219819.3219823}{doi:\nolinkurl{10.1145/3219819.3219823}}


\bibitem[Zhou et~al\mbox{.}(2026)]%
        {zhou2026tokenformer}
\bibfield{author}{\bibinfo{person}{Yifeng Zhou}, \bibinfo{person}{Yuehong Hu},
  \bibinfo{person}{Zhixiang Feng}, \bibinfo{person}{Junwei Pan},
  \bibinfo{person}{Kaihui Wu}, \bibinfo{person}{Hanyong Li},
  \bibinfo{person}{Shangyu Zhang}, \bibinfo{person}{Shudong Huang},
  \bibinfo{person}{Zhangbin Zhu}, \bibinfo{person}{Chengguo Yin},
  \bibinfo{person}{Haijie Gu}, {and} \bibinfo{person}{Jie Jiang}.}
  \bibinfo{year}{2026}\natexlab{}.
\newblock \showarticletitle{{TokenFormer}: Unify the Multi-Field and Sequential
  Recommendation Worlds}.
\newblock \bibinfo{journal}{\emph{arXiv preprint arXiv:2604.13737}}
  (\bibinfo{year}{2026}).
\newblock
\urldef\tempurl%
\url{https://arxiv.org/abs/2604.13737}
\showURL{%
\tempurl}


\bibitem[Zhu et~al\mbox{.}(2025)]%
        {zhu2025rankmixer}
\bibfield{author}{\bibinfo{person}{Jie Zhu}, \bibinfo{person}{Zhifang Fan},
  \bibinfo{person}{Xiaoxie Zhu}, \bibinfo{person}{Yuchen Jiang},
  \bibinfo{person}{Hangyu Wang}, \bibinfo{person}{Xintian Han},
  \bibinfo{person}{Haoran Ding}, \bibinfo{person}{Xinmin Wang},
  \bibinfo{person}{Wenlin Zhao}, \bibinfo{person}{Zhen Gong},
  \bibinfo{person}{Huizhi Yang}, \bibinfo{person}{Zheng Chai},
  \bibinfo{person}{Zhe Chen}, \bibinfo{person}{Yuchao Zheng},
  \bibinfo{person}{Qiwei Chen}, \bibinfo{person}{Feng Zhang},
  \bibinfo{person}{Xun Zhou}, \bibinfo{person}{Peng Xu}, \bibinfo{person}{Xiao
  Yang}, \bibinfo{person}{Di Wu}, {and} \bibinfo{person}{Zuotao Liu}.}
  \bibinfo{year}{2025}\natexlab{}.
\newblock \bibinfo{title}{RankMixer: Scaling Up Ranking Models in Industrial
  Recommenders}.
\newblock
\showeprint[arxiv]{2507.15551}~[cs.IR]
\urldef\tempurl%
\url{https://arxiv.org/abs/2507.15551}
\showURL{%
\tempurl}


\end{thebibliography}

\appendix
\section{Appendix}

\subsection{Related Work}
\label{sec:related_work}

\subsubsection{Multi-Task Recommendation}

Multi-task recommendation jointly models diverse user feedback and business objectives, with a central focus on balancing knowledge sharing and task specialization. Shared-Bottom~\cite{caruana1997multitask} combines a shared encoder with separate prediction towers. MMoE~\cite{ma2018mmoe} introduces a shared expert pool with task-specific gates that adaptively combine expert outputs. PLE~\cite{tang2020ple} organizes shared and task-specific experts into a progressive extraction architecture, while AdaTT~\cite{li2023adatt} enables adaptive knowledge transfer among tasks through gated residual units.

Recent work has extended multi-task recommendation toward model scaling. SMES~\cite{zhang2026smes} expands a sparse expert pool, using task-shared and task-adaptive routing to coordinate expert activation. UniFormer~\cite{chen2025uniformer} integrates feature-space and task-space interactions within a unified architecture, while OneRank~\cite{tang2026onerank} incorporates task-specific modeling at early stages of feature interaction. These developments extend the study of knowledge sharing and task specialization to larger recommendation models. DuoFormer pursues this direction with a focus on expanding fine-grained task-specific modeling capacity alongside shared feature learning.

\subsubsection{Model Scaling for Recommender Systems}

Inspired by the success of model scaling in LLMs~\cite{kaplan2020scaling,hoffmann2022training}, recent research has explored scalable architectures for industrial recommendation. Wukong~\cite{zhang2024wukong} scales model depth and width through stackable factorized interaction layers, while RankMixer~\cite{zhu2025rankmixer} replaces self-attention with efficient token mixing. TokenMixer-Large~\cite{jiang2026tokenmixerlarge} combines mixing-and-reverting operations with improved residual connections and sparse per-token experts to support deeper and larger models. Recent work further explores joint scaling of non-sequential feature interactions and behavior sequence modeling. OneTrans~\cite{zhang2025onetrans} unifies both within a single Transformer~\cite{vaswani2017attention}, while MixFormer~\cite{huang2026mixformer} combines dense-feature mixing with cross-attention over reusable sequence representations. These scaling mechanisms primarily expand the capacity of feature interaction and sequence encoding.

When used as shared backbones followed by task-specific modules for multi-task recommendation, these architectures follow a \emph{sequential paradigm}, where input features are first encoded into shared representations before task specialization is introduced. Scalable multi-task models such as SMES and UniFormer likewise concentrate capacity growth in shared encoding~\cite{zhang2026smes,chen2025uniformer}. Under this organization, early feature interactions remain task-agnostic even as shared capacity grows, leaving downstream task modules to extract task-relevant information from a common, highly abstracted feature view. The granularity of task modeling presents a further limitation. Although OneRank introduces task-specific modeling earlier, it still relies on a single task state and a common feature view across tasks~\cite{tang2026onerank}. Diverse task factors are aggregated without explicitly distinguishing their feature dependencies, and individual feature representations are not explicitly adapted to task-specific requirements.

DuoFormer introduces a \emph{parallel paradigm} for scaling multi-task recommendation, in which fine-grained task modeling proceeds alongside shared feature learning from the earliest stages of feature interaction. It combines structured multi-view task representations with task-specific feature adaptation, enabling increased model capacity to strengthen both shared learning and fine-grained task modeling. Its shared channel reuses feature computation across tasks, while lightweight task-specific adaptation limits the additional overhead of specialization, supporting efficient scaling.

\subsection{Visualization Analysis}
\begin{figure}[t]
\centering
\includegraphics[width=1.0\linewidth]{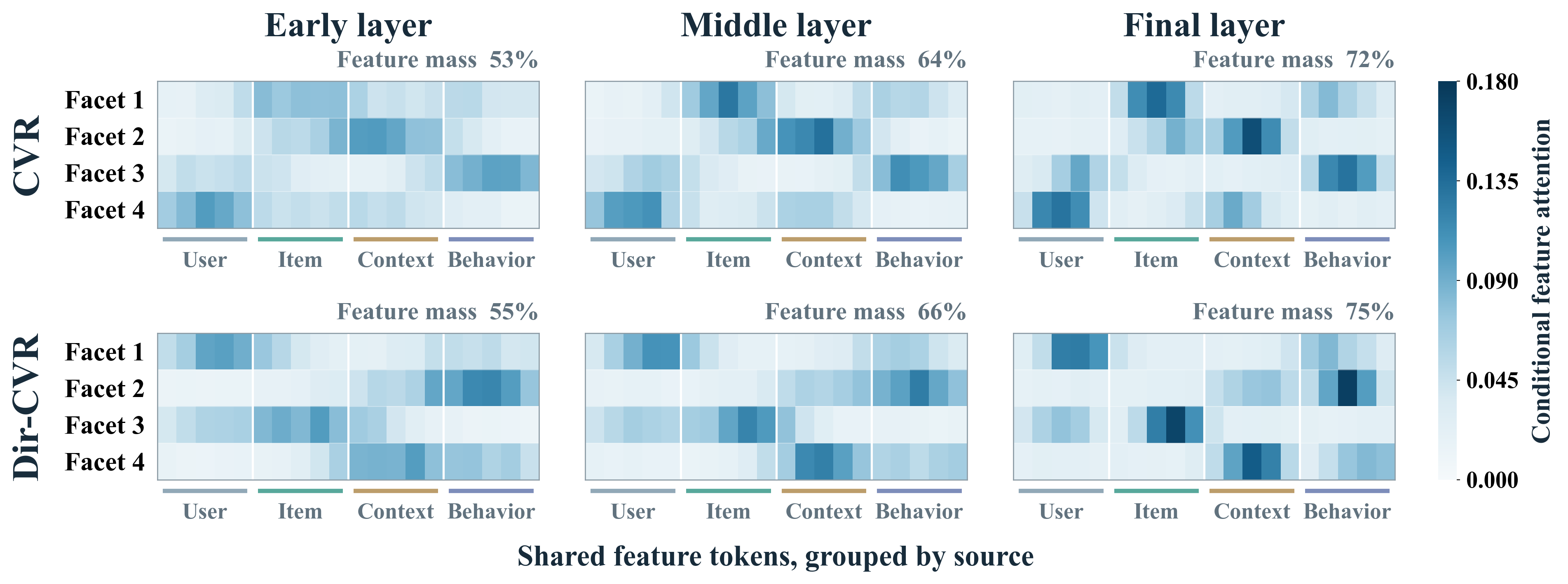}
\caption{Visualization of facet-to-feature attention across tasks and interaction layers.}
\label{fig:facet_attention}
\end{figure}

We use visualizations to examine two questions central to DuoFormer's design: (1) \emph{Do distinct task views emerge early and become progressively refined during shared feature learning?} (2) \emph{How does the low-rank channel adapt the same shared feature to different tasks in terms of relevance assessment and information representation?}

\subsubsection{Early Multi-View Task Specialization}

Figure~\ref{fig:facet_attention} visualizes facet-to-feature attention across interaction layers. Colors show each facet's relative attention to shared feature tokens, with each row normalized to sum to one. Feature mass reports the mean total attention assigned to shared features in the original distribution over all permitted tokens.

For both CVR and Dir-CVR, facets attend to different subsets of user, item, context, and behavior features, showing that each task retrieves evidence through multiple differentiated views. These preferences are already distinct in early layers, indicating that task-specific retrieval begins during the early stages of shared encoding. With increasing depth, the preferences become more concentrated while remaining distinct across facets, illustrating progressive refinement of multiple task views. Meanwhile, the increasing feature mass shows that facets allocate a growing proportion of their total attention to shared features. Together, the early emergence of distinct retrieval patterns and their continued refinement across layers support our motivation to develop fine-grained task representations alongside shared feature learning. Maintaining separate facets allows each task to refine multiple feature views throughout the backbone, providing diverse evidence for task-level integration.

\begin{figure}[t]
\centering
\includegraphics[width=1.0\linewidth]{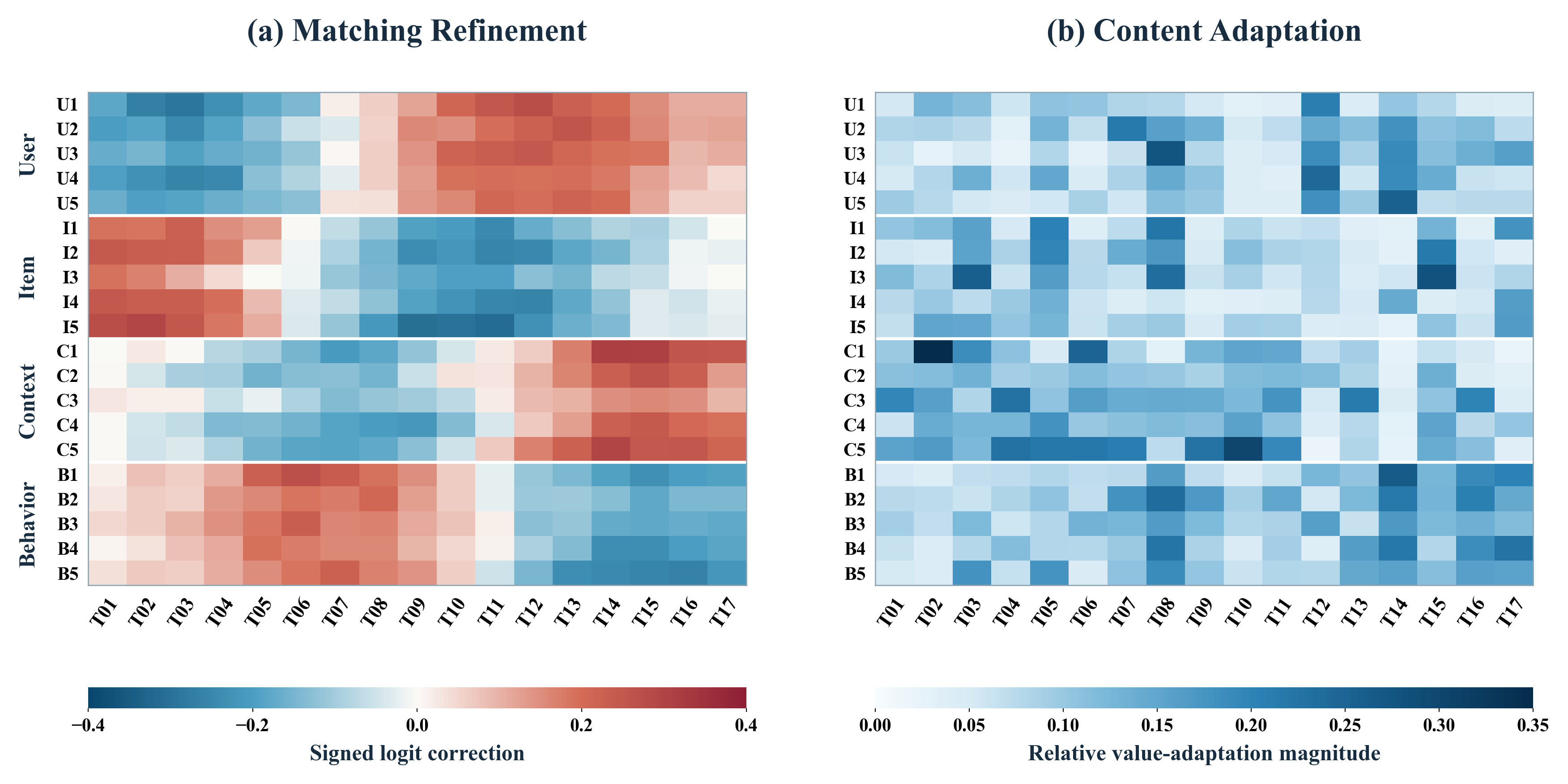}
\caption{Visualization of task-specific feature adaptation across tasks. (a) Signed corrections to feature-matching logits. (b) Relative magnitudes of value adaptations.}
\label{fig:feature_interpretation}
\end{figure}

\subsubsection{Task-Specific Feature Interpretation}

Figure~\ref{fig:feature_interpretation} visualizes how the low-rank task-specific channel adapts the same shared features to different task requirements. Panel~(a) shows signed corrections to feature-matching logits induced by key adaptations, while panel~(b) reports value-adaptation magnitudes relative to the shared value representations.

The same feature receives positive matching corrections for some tasks and negative corrections for others, illustrating how feature adaptation adjusts its relevance to different objectives. Value-adaptation magnitudes also vary across tasks for the same feature, indicating that its information representation is refined to different extents according to task requirements. These patterns support our motivation to adapt the feature representations themselves, allowing individual tasks to evaluate feature relevance and interpret feature information according to their own requirements. Complementing the feature-adaptation ablation, this visualization illustrates how the low-rank channel modifies both matching scores and value representations for individual tasks while preserving a reusable shared feature basis.
\end{document}